%% file: arXiv_main.tex
\documentclass[11pt]{article}
\usepackage[margin=1in]{geometry}
\usepackage[utf8]{inputenc}
\usepackage{amsmath,amssymb,amsfonts}
\usepackage{graphicx}
\usepackage{booktabs, multirow}
\usepackage{algorithmic}
\usepackage{subcaption}
\usepackage{float}
\usepackage{balance}
\usepackage{tcolorbox}
\usepackage[numbers,sort&compress]{natbib}
\usepackage[font=small,labelfont=bf]{caption}
\usepackage{url}
\usepackage{xcolor}
\usepackage{color,soul,xspace}
\usepackage[colorlinks,linkcolor=blue,citecolor=blue,urlcolor=blue]{hyperref}
\graphicspath{{figs/}{figures/}{images/}}   
\DeclareGraphicsExtensions{.pdf,.png,.jpg}  

\usepackage{newunicodechar}
\newunicodechar{≥}{\geq}
\newunicodechar{μ}{\mu}

\newcommand{\vsa}{\vspace*{-0.28cm}}

\newcommand{\mourmeth}{\text{PrIVAE}}
\newcommand{\ourmeth}{$\mourmeth$\xspace}

\title{Property-Isometric Variational Autoencoders for Sequence Modeling and Design}
\author{
  Elham Sadeghi$^{1}$ \and
  Xianqi Deng$^{1}$ \and
  I-Hsin Lin$^{2}$ \and
  Stacy M. Copp$^{2}$ \and
  Petko Bogdanov$^{1}$
}

\date{
$^{1}$Department of Computer Science, University at Albany- SUNY, Albany, NY, USA\\
$^{2}$University of California- Irvine, Irvine, CA, USA\\[1ex]
}

\begin{document}
\maketitle

\begin{abstract}
Biological sequence design (DNA, RNA or peptides) with desired functional properties has applications in discovering novel nanomaterials, biosensors, anti-microbial drugs and beyond. One common challenge is the ability to optimize complex high-dimensional properties such as target emission spectra of DNA-mediated fluorescent nanoparticles, photo and chemical stability, and antimicrobial activity of peptides across target microbes. Existing models rely on simple binary labels (e.g., binding/non-binding) as opposed to high-dimensional complex properties. To address this gap, we propose a geometry-preserving variational autoencoder framework, called \ourmeth, which learns latent sequence embeddings that respect the geometry of their property space. Specifically, we model the property space as a high-dimensional manifold that can be locally approximated by a nearest neighbor graph, given an appropriately defined distance measure. We employ the property graph to guide the sequence latent representations as 1) GNN encoder layer(s) and 2) an isometric regularizer. \ourmeth learns a property-organized latent space that allows rational design of new sequences with desired properties by employing the trained decoder. We evaluate the utility of our framework for two generative tasks: 1) design of DNA sequences that template fluorescent metal nanoclusters and 2) design of anti-microbial peptides. The trained models retain high reconstruction accuracy while organizing the latent space according to properties. Beyond in silico experiments, we also employ sampled sequences for wet lab design of DNA nanoclusters resulting in up to 16.1-fold enrichment of rare-property nanoclusters compared to their abundance in training data and demonstrating the practical utility of our framework.
\end{abstract}


\section{Introduction}
\begin{figure*}[ht]
    \centering
    \includegraphics[width=0.75\textwidth, trim=1.3cm 8.4cm 0.5cm 0pt, clip]{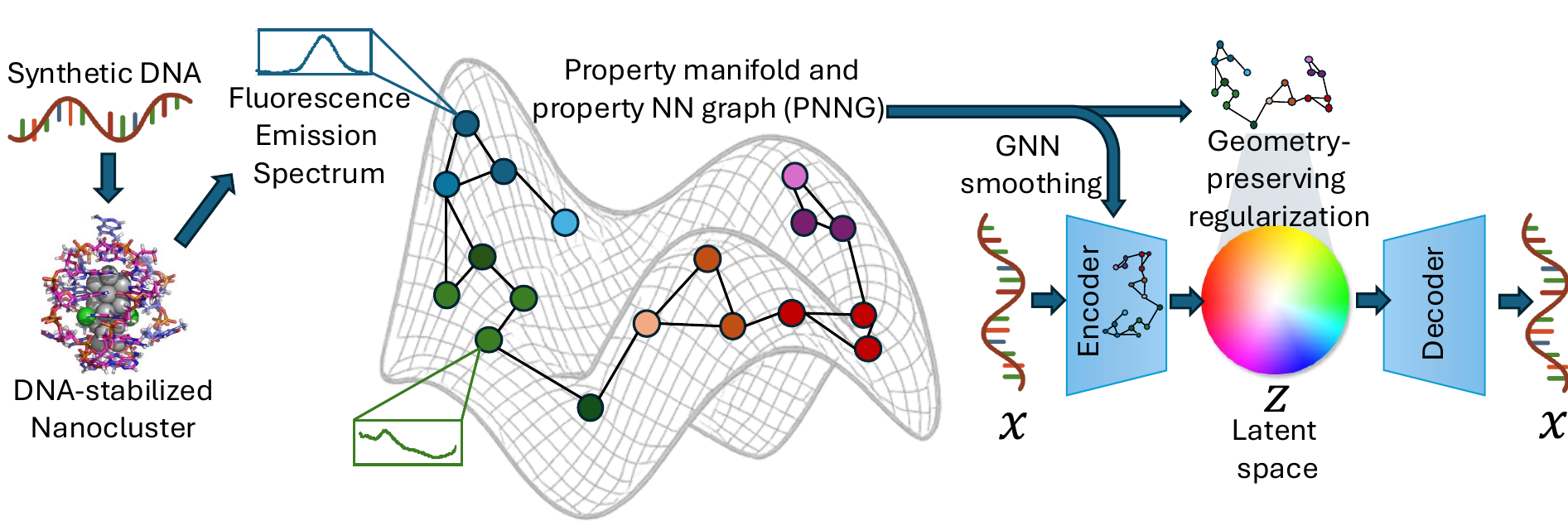}\vsa
    \caption{\footnotesize Overview of the proposed geometry-preserving framework. We hypothesize that complex properties of biological sequences (e.g., emission spectra of DNA nanoclusters or peptide antimicrobial activity) lie on a manifold which we approximate locally by a property nearest neighbor graph (PNNG). The PNNG is utilized in two ways in our autoencoder framework: as GNN layers in the encoder which smooth sequence representations based on similar-property neighbors; and for isometric regularization of the latent space. The property-organized latent space allows for simple generation by sampling in target property enriched regions and decoding to obtain candidate sequences.
    }\vsa
    \label{fig:Schematic}
\end{figure*}

Biological sequences are programmable building blocks for both living organisms but also in engineered systems like novel biomaterials, drugs and in synthetic biology~\cite{heinemann2006synthetic,holmes2002novel}.As a result, designing newly engineered synthetic biological sequences with desired functional properties is a central challenge in many applications~\citep{zhang2024artificial,mardikoraem2023generative,copp2019general}. Depending on the target application, the sequences and the products they enable may exhibit complex, high-dimensional properties tuned by the sequence. For example, DNA-stabilized metal nanoclusters composed of small metal cores and stabilized by single-strand synthetic DNA sequences (Fig~\ref{fig:Schematic} Left) fluoresce when excited across the visible and NIR spectrum. Tuning this emission spectrum by the stabilizing DNA sequence is important for creating new deep-tissue biosensors~\citep{gonzalez2024atom, moomtaheen2022dna,guha2023electron}. Similarly, peptide sequences can be engineered to have antimicrobial activity profiles against a number of different bacteria~\citep{szymczak2023discovering}. Optimizing complex properties like emission spectra and antimicrobial profiles requires generative models handling more than binary labels and accounting for the geometric structure of the property space~\citep{greener2018design,mastracco2022chemistry,barazandeh2024utrgan}.

Current generative models for biological sequence design offer limited controllability, as they are typically conditioned on simple binary or scalar labels such as binding vs non-binding or high vs low activity~\citep{gupta2018feedback, yu2023multi,moomtaheen2022dna}. This oversimplification fails to capture the continuous and multi-dimensional nature of biological properties and shifts the burden to downstream classifiers, which often overfit small datasets and perform poorly on out-of-distribution sequences. Recent diffusion-based approaches~\cite{soares2025targeted,torres2025generative,jin2025ampgen} and GFlowNets~\citep{jain2022biological,jain2023multi} extend generative capabilities, but remain constrained to binary or simplified objectives and rely on accurate oracles that are unavailable for many complex biological phenomena. These limitations are particularly pronounced in synthetic biological settings, where experimental assays are costly and complex property profiles, such as DNA nanocluster spectra or multi-bacteria antimicrobial activity, are rare and sparsely represented. Our goal in this work is to overcome these challenges by enabling rational design of biological sequences with multi-objective, high-dimensional properties, even in limited-data regimes.

We introduce a geometry-preserving variational autoencoder (VAE) framework, called \ourmeth, which aligns the sequence embedding space with the geometry of associated properties (Fig.~\ref{fig:Schematic}). Central to \ourmeth is the hypothesis that biosequence properties like fluorescence spectra belong to a high-dimensional manifold which can be locally approximated by a property nearest neighbor graph (PNNG). We learn a latent space that preserves the property geometry and can be used for designing new sequences. To this end, we utilize the PNNG in two ways. First, we incorporate GNN layers in the encoder to ``align'' sequence representations to those of neighbors with similar properties. Second, we employ an isometric regularization of the latent space $z$ that is also designed to ensure that sequences with similar properties have similar representations. The resulting property-organized latent space allows for simple sequence generation by sampling in regions enriched in target properties and decoding to obtain candidates. We demonstrate the effectiveness of our framework for the design of DNA-stabilized nanoclusters and antimicrobial peptides. Our models maintain high reconstruction accuracy while organizing the latent space according to properties, thus enabling simple interpretable property-aware generation. Beyond \emph{in silico} experiments, we also verify designed DNA-templated nanoclusters in the \emph{wet lab}, enriching rare-property nanoclusters by $16.1$ folds compared to their abundance in training data.

Our contributions in this work are as follows:\\
$\bullet$ We introduce a geometry-preserving generative framework, called \ourmeth, that organizes the latent space based on the structure of complex, high-dimensional sequence properties.\\
$\bullet$ We demonstrate the utility of \ourmeth in two practical applications: fluorescent Ag\textsubscript{N}-DNAs and antimicrobial peptides; both achieving controllable, property-aware generation.\\
$\bullet$ We validate \ourmeth's predictions by both \emph{in silico} benchmarking and experimental wet lab synthesis of novel DNA-stabilized nanoclusters with desired emission characteristics.

\section{Related work}

\noindent\textbf{Biological sequence design.}
A broad range of generative models have been employed to synthesize biological sequences~\cite{zhu2024generative}. Early work leveraged Variational autoencoders (VAEs) to map sequences into latent spaces and back, enabling protein variant discovery, aptamer and genome synthesis, and expression tuning~\citep{shcherbakova2024designing, praljak2023protwave}. Generative adversarial networks (GANs) were also used to explore the functional protein space, modulate gene expression, generate 5'UTRs and RNA-protein interactors and augment RNA datasets~\citep{barazandeh2024utrgan, ozden2023rnagen, chiquitto2024generative}.
More recently transformer architectures, pretrained on large protein or RNA corpora or fine-tuned on SELEX datasets, have emerged as reliable tools for aptamer design and family-agnostic protein generation~\citep{andress2023daptev, zhang2024rnagenesis, madani2023large}. Other generative approaches like diffusion models, flow-matching methods, and Generative Flow Networks (GFlowNets) have also been explored for biological sequence and molecule design, enabling trajectory-based generation or phenotype-conditioned editing~\citep{li2024discdiff, huang2024latent, dasilva2024dna,lu2024cell}. In particular, GFlowNets have been applied to AMP generation where sequences are sampled proportional to an oracle predicting binary AMP activity ~\cite{jain2022biological}, and later extended to multi-objective settings with properties such as binding affinity, toxicity, and solubility through preference-conditioned or Pareto-optimal sampling~\cite{jain2023multi}.

Despite the diversity of these generative approaches, from VAEs, GANs, and transformers to newer models like diffusion and flow-matching, most of them share common limitations. These methods typically rely on large, property-diverse datasets, curated annotations, or pretraining on natural sequences, and optimize simple properties (active/inactive). In the case of GFlowNets, existing applications remain constrained to binary labels in AMPs, and fundamentally depend on reliable oracles. No such robust oracle exists for complex biological properties, especially in DNA nanoclusters where properties are rare and complicated, which makes direct application of GFlowNets infeasible in our setting. Such assumptions limit the generalizability of current methods in synthetic biological contexts, where experimental data is not as abundant since assays are costly, and properties arise from complex, experimentally measured phenomena that cannot be simply represented as scalar or binary labels.

\noindent\textbf{Conditioning on properties.}
Generative modeling has frequently been adapted to incorporate conditioning on desired biological sequence properties. Many approaches, particularly in Antimicrobial Peptide (AMP) design, condition on relatively simple labels such as post-hoc predicted scalar values (e.g., MIC), or direct binary/continuous activity labels, restricted to single-pathogen or single-attribute training data~\citep{dean2020variational, dean2021pepvae, szymczak2023discovering, yu2023multi, das2018pepcvae}. These methods fail to capture complex high-dimensional functional nuances. Some methods have targeted multi-objective optimization or property-aware sampling~\citep{praljak2023protwave, ozden2023rnagen, surana2023pandoragan}. Yet, even when tackling inherently complex experimental properties like DNA-stabilized silver nanocluster spectra, models frequently resort to simplifications, such as optimizing for only a single spectral peak~\citep{moomtaheen2022dna, sadeghi2024multi}. Recent advances in AMP generation with diffusion models have explored classifier-guided and conditional strategies, including OmegAMP (species- and strain-specific classifiers), AMP-Diffusion (conditioning on general potency and toxicity in ESM-2 space), and AMPGen (evolutionary information through multiple sequence alignment conditioning with post hoc MIC scoring). Despite methodological differences, all three approaches share key limitations: reliance on binary activity or toxicity labels, lack of robustness to rare species, and inability to condition generation on multiple species simultaneously~\cite{soares2025targeted,torres2025generative,jin2025ampgen}.

Since existing property-conditioned methods utilize simplified property labels, regularize isolated latent dimensions, or assume properties are ordered and decomposable along orthogonal axes, they are ill-suited for complex, interdependent properties like emission spectra or multi-bacteria antimicrobial profiles. These challenges, coupled with sparse and biased annotations in many design problems present a critical limitation: a general inability to effectively leverage and model the rich and nuanced underlying structure of high-dimensional property spaces when designing new sequences.

\noindent\textbf{Foundational models.}
Recent efforts in foundational modeling for biological sequences leverage large-scale pretraining on natural biological sequence corpora (e.g., Bio-xLSTM~\citep{schmidinger2024bio}) to achieve generalization across biological domains, often supporting tasks like in-context learning. Our framework is complementary to such approaches. For instance, pretrained embeddings from such foundational models can serve as input representations for our methods, potentially enhancing its performance for sequences that share characteristics with natural ones. Indeed, this is a strategy we adopt for antimicrobial peptides design in this work resulting in better results than employing one-hot encoding. However, such an approach has limitations for applications relying on synthetic sequences where functional motifs and sequence-property relationships are fundamentally different from those shaped by natural evolution. For example, in Ag\textsubscript{N}-DNAs, the photo-physical properties emerge from complex interactions between DNA bases, the silver nanocluster core, and solution chemistry. As a result, transfer learning from natural sequences has limited applications and custom embedding methods trained on carefully curated experimental datasets are necessary for property-controllable sequence generation.

\noindent\textbf{Geometry-preserving modeling.}
Isometric embedding models aim to preserve the intrinsic data geometry in latent representations, especially when functional supervision is weak. GRAE~\cite{duque2022geometry} aligns the latent space with a reference embedding that preserves local and global relationships in the input data. IRVAE~\cite{lee2022regularized} and FlatVI~\cite{palmaenforcing} regularize the decoder’s Jacobian or Fisher information pullback to impose local isometry or uniform latent geometry, based on distances in the data space.
A separate line of work incorporates predefined structural graphs to guide representation learning. Xu et al.\cite{xu2024beyond} and Krapp et al.\cite{krapp2024context} use base-pairing graphs or atomic point clouds derived from experimentally determined RNA and protein structures, while GGNN~\cite{zhu2023geometric} operates on curated biological interaction networks such as protein–protein interaction graphs. In these settings, the graph encodes known physical or biochemical relationships and is fixed prior to training.
In contrast, we construct a graph from sequence properties (as opposed to the sequences themselves) and aim to preserve the geometry of the property space in the latent sequence representations. 

\section{Problem formulation}
Our goal is to design new biological sequences represented as one-hot encodings \( X\in \{0,1\}^{l\times a} \), where $l$ is the sequence length and $a$ is the size of the alphabet. Designed sequences should exhibit high-dimensional functional properties of interest \( y \in \Re^{p}\). Importantly, while properties are represented by high-dimensional vectors, not all property configurations are possible due to physical constraints. For example, fluorescence spectra of DNA-stabilized nanoclusters are concentrated around wavelength peaks corresponding to the sizes of particles they stabilize. Similarly, due to their geometry when folded, peptides may be active against some groups of bacteria and not against others. Since the constraints are not known in advance, we impose a manifold assumption on the properties, namely properties are sampled from a smooth lower-dimensional manifold $\mathcal{M}$. We also assume that there exists a domain-appropriate distance measure $d(y_i,y_j)\mapsto \Re$ that quantifies the difference between the properties of a pair of sequences. The input training dataset comprises $n$ pairs of samples $\mathcal{D}=\{(X_i, y_i)\}$.

\begin{tcolorbox}[colback=gray!5!white,colframe=black!75!black]
  \noindent{\bf Problem: \emph{Property-based sequence generation}}. Given a dataset $\mathcal{D}$, train a generative model that can be employed to design new sequences $X^{*}$ based on a target desired property configuration $y^{*}$. 
\end{tcolorbox}
\section{Method: Property Isometric VAE (PrIVAE)}
Just like natural biosequences, synthetic bio-sequences contain distinct subsequence motifs determining their properties. Our goal is to model them via low-dimensional latent embeddings that jointly capture essential motifs and ``respect'' the geometry of the property space. Intuitively sequences with similar properties should have similar embeddings. To this end, we propose a \emph{Property Isometric Variational Autoencoder (PrIVAE)} framework which in addition to the usual reconstructive VAE objective aims to approximate the property manifold in latent space using two mechanisms: (i) an isometric regularizer aligning the latent space with the property geometry, and (2) graph neural network layers structured according to the property similarity graph. We discuss the individual components as well as batch training in the rest of this section.

\subsection{Preliminaries: basic VAE for sequences}
\begin{figure}[t]
    \centering
    \includegraphics[width=\linewidth, trim=110 90 110 90, clip]{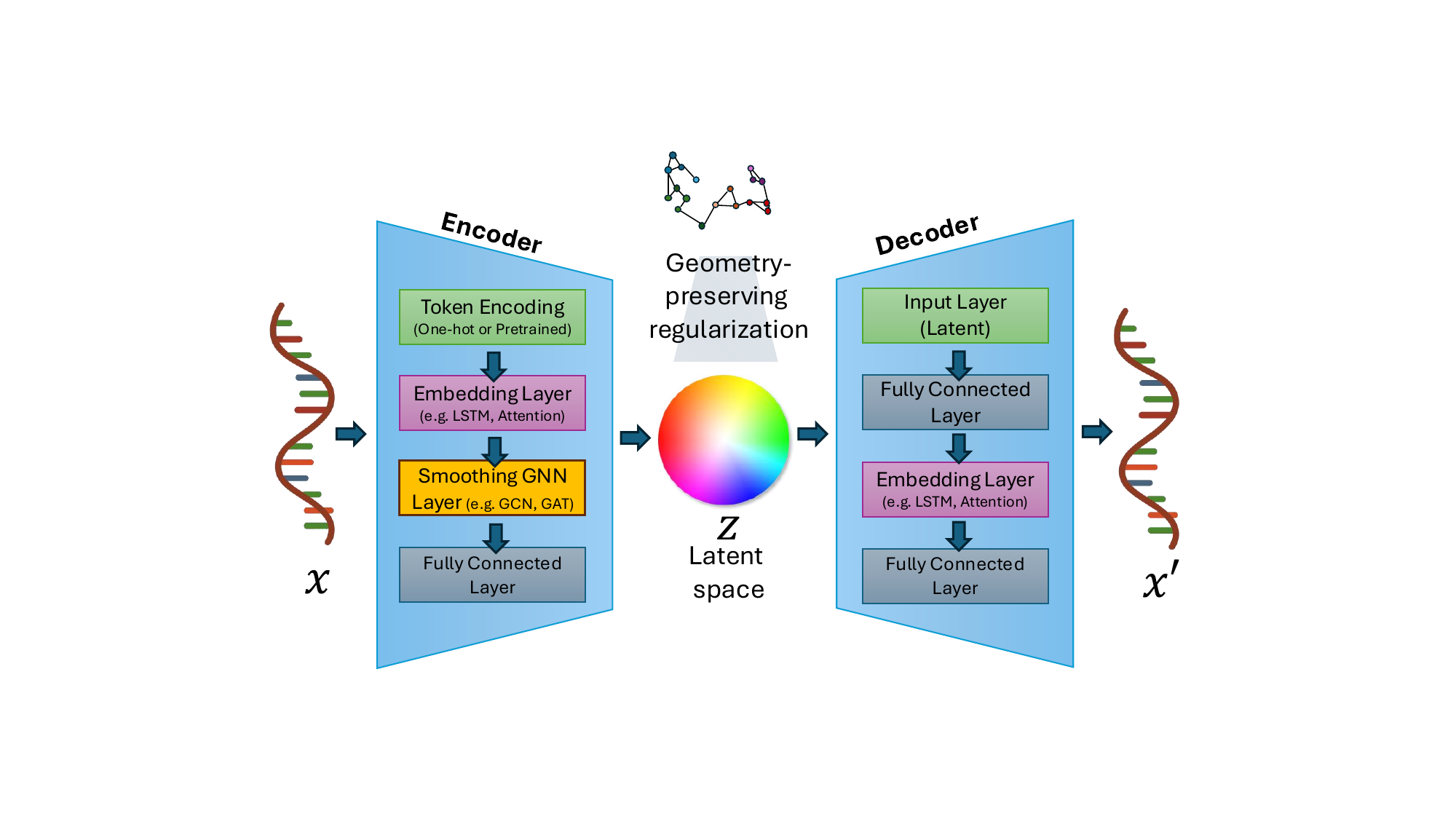}
    \caption{\footnotesize End-to-end architecture of our proposed VAE model.}
    \label{fig:arch}\vsa
\end{figure}

We first describe the basic architecture that allows ``unconditional'' embedding and sampling of sequences without considering their properties. Our goal is to both model low-dimensional signals from  the sequences (e.g., motifs) and sample new sequences from a low-dimensional latent space. Hence we resort to an encoder-decoder architecture. Specifically, we employ a $\beta$-VAE~\cite{higgins2017beta}, where the encoder maps an input sequence \( X_i \) to a posterior distribution \( q_\phi(z_i|X_i) \) in latent space, and the decoder reconstructs the sequence via \( p_\theta(X_i|z_i) \).
The training objective is to minimize the negative ELBO, written as:

\begin{equation}
\mathcal{L}_{\text{VAE}} = -\,\mathbb{E}_{q_\phi(z|X)}[\log p_\theta(X|z)] 
+ \beta \cdot \text{KL}\!\left(q_\phi(z|X) \,\|\, p(z)\right)
\end{equation}

where \( p(z) = \mathcal{N}(0, I) \) is an isotropic Gaussian prior distribution. The first term promotes accurate reconstruction, while the second term encourages independence among the latent dimensions by ``aligning'' the posterior distribution to a normal prior controlled by a hyperparameter \( \beta \).
The specific architecture of the VAE we adopt is presented in Fig~\ref{fig:arch} and described below. Note, that the latent representation $z_i$ corresponds to the whole sequence $X_i$ as opposed to individual tokens.

\noindent{\bf Input sequence encoding.} Depending on the input data domain, we have different choices of input token representation (a token in our setting is a single sequence character such as a nucleotide base or amino acid). If working with natural sequences like peptides and their anti-microbial activity, we expect that important sequence motifs will be shared with those of natural proteins, and hence we can employ pre-trained token representation models like ESM-2~\citep{lin2023evolutionary}. Alternatively, when working with synthetic sequences like singl-strand DNAs that template nanoclusters, we employ one-hot encoding to represent the individual tokens. 

\noindent{\bf Encoder/decoder architecture.} Since our input modality is sequences, natural choices for the encoder and decoder are recurrent or attention-based layers. We experiment with both bi-directional Long short-term memomory layers (LSTM) and attention layers to encode and decode sequences. In each case the sequence representation layer(s) are followed by a fully connected layer. Details of tuning the number of layers, their sizes and other architectural hyperparameters are available in the Appendix. Where our architecture differs from traditional VAEs is in the additional smoothing GNN layer following the encoder and our overall training objective described next. 

\subsection{Learning a property isometric latent space}

The basic VAE framework does not impose structure on the latent space apart from independence among latent variables through the KL-divergence term. Our goal is to learn low-dimensional embeddings for the whole sequences $X$ that are also ``aligned'' with high-dimensional properties $y$ sampled from a smooth manifold where local geometry reflects functional similarity. Intuitively, we want to preserve the property $y$ geometry in the learned latent space $z$. Exact isometry preservation including all pairwise distances and angles is typically infeasible in high-dimensional settings. A typical approach in manifold learning is to preserve local neighborhoods that are sufficient to recover useful low-dimensional structure~\citep{tenenbaum2000global,roweis2000nonlinear}. Under the smoothness assumptions of the property manifold, the tangent space of a manifold can be locally approximated using nearest neighbors~\citep{belkin2003laplacian}, making locality-preserving methods particularly effective.

To enforce alignment between latent representations and the property manifold, we employ two strategies: \emph{isometric regularization} and \emph{graph-based smoothing}. Both strategies rely on a \emph{Property Nearest Neighbor Graph (PNNG)} which we construct based on the properties $y_i$ of training instances.

\noindent{\bf PNNG construction.}
We follow a standard paradigm in isometric learning. That is, we aim to learn embeddings that ``respect'' nearest neighbor distances since locally the smooth manifold resembles a Euclidean space~\citep{belkin2003laplacian}. To this end we construct a weighted $k$ nearest neighbor graph $G_y(V,E,W)$, called Property Nearest Neighbor Graph (PNNG), with node set $V$, edge set $E$ and weighted adjacency matrix $W$. The weights on edges are based on pair-wise distances \( d(y_i,y_j \). The employed distance metric is domain-specific and should measure a meaningful distance depending on the semantics of the properties $y$. For example, the emission spectra in our DNA nanocluster dataset are mixtures of peaks with height, width and mean parameters (akin to Gaussian mixtures though not 1-normalized). Since $L_p$ distance measures are not appropriate for properties represented as sets of peaks, we employ a Cauchy-Schwarz divergence measure~\citep{kampa2011closed}. Alternatively, when computing the PNNG for the antimicrobial activity of peptides, we employ the Manhattan ($L_1$) distance over Minimum Inhibitory Concentration (MIC) profiles. Definitions of these distances are available in Appendix~\ref{append:dist}. The edge weights which have the semantics of similarity in the PNNG graph are computed by transforming the distances using the radial basis function (RBF) kernel, defined as \( w_{ij} = \exp(-d(y_i,y_j)^2 / 2\sigma^2) \), where \( \sigma \) controls the sensitivity of the kernel. 

\noindent{\bf Isometric regularization.}
To align the latent space $z$ with the property geometry approximated by the PNNG $G_y$, we add a regularization term based on Laplacian Eigenmaps~\citep{belkin2003laplacian} similar to other locality-preserving methods~\citep{meilua2024manifold,cayton2008algorithms}. Let \( Z = \{z_1, \dots, z_n\} \) be the latent embeddings of training data instances, and \( W \in \mathbb{R}^{n \times n} \) be the PNNG adjacency matrix. Let also $L$ denote a Laplacian matrix associated with $G_y$. We then define an isometric regularization term as follows:
\begin{equation}
\mathcal{L}_{\text{ISO}} = \text{Tr}(Z^\top L Z),
\end{equation}
where $\text{Tr}()$ denotes the trace of the the matrix operand. Intuitively, this term penalizes embeddings which have high similarity in property space (i.e., nearby on the property manifold) and low similarity in latent space. In terms of Laplacian matrices, we have a choice among the combinatorial, normalized, random walk Laplacians and others~\cite{meilua2024manifold}. In our experiments, we adopted the combinatorial Laplacian which performed well in the two target domains we considered, though alternatives should be considered when employing the method for other tasks. 

For completeness, we define the combinatorial Laplacian below and the explicit form of our isometric regularization loss. Let \( D \) be the degree matrix of $G_y$ with diagonal entries \( D_{ii} = \sum_j W_{ij} \). The combinatorial graph Laplacian is defined as \( L = D - W \) and the additional isometric loss term is:
\begin{equation}
\mathcal{L}_{\text{ISO}} = \text{Tr}(Z^\top L Z) = \frac{1}{2} \sum_{i,j} W_{ij} \| z_i - z_j \|_2^2.
\end{equation}
Minimizing this loss encourages neighboring sequences in the property space to remain close in latent space, thus approximating local isometry. 
It is important to note that unlike the majority of the isometric learning literature which aims to preserve the geometry of the \emph{input space}~\citep{lee2022regularized,lim2024graph}, our approach targets the \emph{property space}, where distances capture functional similarity. This distinction makes it effective for controllable sequence design involving high-dimensional complex properties.

\noindent{\bf Graph-based smoothing.}
Apart from the regularization, we also employ a ``more direct'' approach to align representations \( h_i \in \mathbb{R}^h \) of PNNG neighbors based on Graph Neural Networks (GNNs) where nodes learn to aggregate information from similar property neighbors~\citep{gilmer2017neural}. 
GNN architectures differ in how they weigh and aggregate neighbor information. We experimented with several options, including Graph Convolutional Networks (GCN)~\citep{kipf2016semi} and Graph Attention Networks (GAT)~\citep{velivckovic2017graph} and adopted the former in our experiments due to its better empirical performance in our setting. The smoothed hidden representation in GCNs is computed as:

\begin{equation}
h'_i =  \sum_{j \in \mathcal{N}(i)} \frac{w_{ij}}{\sqrt{w_i w_j}} \Theta h_j ,
\end{equation}
where $h_i$ and $h'_i$ are the hidden representations of node $i$ before and after the current GCN layer, \( w_{ij} \) is the PNNG similarity edge weight between nodes $i$ and $j$, \( w_i=\sum_jw_{ij} \) is the weighted node degree and \( \Theta \) is a learnable weight matrix. This refinement injects contextual information from functionally similar sequences into the encoded representation before computing the final latent representation \( z \).

\noindent{\bf Overall objective and batching.}
The overall training loss combines the VAE objective with isometric regularization:

\begin{equation}
\mathcal{L}_{\text{Total}} = \mathcal{L}_{\text{REC}} + \beta \cdot \mathcal{L}_{\text{KL}} + \gamma \cdot \mathcal{L}_{\text{ISO}},
\end{equation}
where $\mathcal{L}_{\text{REC}}$ is the reconstruction loss, $\mathcal{L}_{\text{KL}}$ is the KL divergence loss, and $\mathcal{L}_{\text{ISO}}$ enforces alignment between latent and property geometries. The hyperparameters $\beta$ and $\gamma$ balance the contributions of each term.

Notably, the two geometry-preserving mechanisms, graph-based smoothing and isometric regularization, operate jointly in an end-to-end manner rather than sequentially or post-hoc. 
The sequence encoder (LSTM or attention) first produces hidden representations $h_i$, which are refined by GNN smoothing over the fixed PNNG. The refined vectors $h'_i$ are then used to parameterize the approximate posterior $q_\phi(z_i \mid X_i)$. Concurrently, the isometric regularization term $L_{\text{ISO}}$ is applied directly to the posterior means of the latent embeddings $z_i$ as part of the total loss. 
Gradients from both the VAE reconstruction/KL objective and the isometric regularizer propagate through the entire network. 
As a result, the sequence representations are shaped from the outset by both reconstruction fidelity and adherence to the geometry of the property manifold. This design guarantees that the encoder, GNN smoothing, and latent regularization are learned jointly, not in alternating stages.

Since we have GNN layers, when creating minibatches for training we need to ensure that they form connected subgraphs involving sequences of similar properties. To this end, we create ``core'' minibatches based on the properties $y$ and augment each minibatch with immediate one-hop neighbors whose hidden representations are employed in the GNN layers only to update the core batch members. Details of the core minibatch creation are available in Appendix~\ref{append:param}.

\section{Experimental evaluation}
We experiment with two datasets: DNAs that template fluorescent silver nanoclusters (Ag\textsubscript{N}-DNAs)~\citep{copp2019general,mastracco2022chemistry} and antimicrobial peptides~\citep{witten2019deep}. We investigate the training performance, latent space organization and the quality of newly designed sequences. We quantify reconstruction \emph{Accuracy} as the fraction of correctly reconstructed sequence symbols: 
\begin{equation}
ACC=1 - \frac{d_H(S, S')}{l},
\end{equation}
where \(d_H\) is the Hamming distance between the original sequence \(S\) and its reconstruction \(S'\).

To quantify the latent space organization we employ a measure of \emph{Purity} profiling the similarity of properties of neighbors in latent space. For a training (or testing) instance $z_i$ in latent space, it is defined as:
\begin{equation}
\text{Purity}_k(i) = \frac{1}{k} \sum_{j \in \mathcal{N}_k(i)} \frac{|C_i \cap C_j|}{|C_i \cup C_j|},
\end{equation}
where \( \mathcal{N}_k(i) \) denotes the \( k \)-nearest neighbors of sequence \( z_i \) in latent space, and \( C_i \) is a sets of property ``landmarks'' or pseudo-labels associated with the property vector of \(y_i\). A landmark can be a spectral peak type of an (Ag\textsubscript{N}-DNAs) or sufficiently strong (based on a threshold) antimicrobial activity against a specific bacteria for peptides. We resort to this definition as it is more meaningful for downstream tasks than quantifying distances in property space $y$. We define the property pseudo-labels for the two datasets in the following sections. We employ the average \( \text{Purity}_{15} \) as a latent space organization metric to tune and characterize trained models. While alternative $k\neq 15$ could be considered, 15 was the smallest value we grid-searched for PNNG creation (we tuned the PNNG neighborhood size for each model, details in the Appendix.) In addition, $15$ does not exceed the frequency of the rarest property configurations for both experimental datasets. 

We split both datasets randomly into training (85\%) and validation (15\%) subsets to fine-tune and measure model performance. When we eventually do \emph{de novo} sequence design, we employ the full datasets to train the generative model used for sampling. To evaluate newly generated sequences, we perform wet lab synthesis for DNA nanoclusters and employ a predictive oracle for antimicrobial peptides. We also perform ablation studies to examine the impact of model components, including graph-based smoothing and isometric regularization.

\subsection{Ag\textsubscript{N}-DNA design}
\noindent{\bf Data.}  
The Ag\textsubscript{N}-DNA dataset consists of $3257$ single-strand DNA sequences of length \( l = 10 \) associated with experimentally measured emission spectra $y$ of the silver nanoclusters they stabilize~\cite{sadeghi2024multi}. Each spectrum consists of up to $4$ peaks characterized by three parameters: i) central wavelength $\lambda$, ii) intensity weight $v$ characterizing the brightness of the peak, and iii) peak width (or spread) $\sigma$. To reduce the dynamic range of intensities, the original weights are transformed using a \(\log_{10}\) normalization.

Since the sequence dictates the spectral properties of Ag\textsubscript{N}-DNAs, the design goal is to predict new sequences with bright peaks in specific wavelength ranges which can be used for biosensors and in other biophotonic applications~\cite{sadeghi2024multi}. While our model employs the continuous representation of the properties $y$, to quantify purity in the latent space and for visualization purposes we adopt spectral landmark regions based on the color ranges defined in~\citep{copp2021large}: Green (G, 400–590 nm), Red (R, 590–660 nm), Far-Red (F, 660–800 nm), and Near-Infrared (N, \( > 800 \) nm). Each sequence is labeled with up to $4$ peaks (pseudo labels) in descending order of their intensity $v$. For example, GN is the label of a sequence with a green brightest peak and a near-infrared second brightest peak.  

\begin{figure}[t]
    \centering

    \begin{subfigure}{0.48\columnwidth} 
        \includegraphics[width=\linewidth,trim=0 350 0 350,clip]{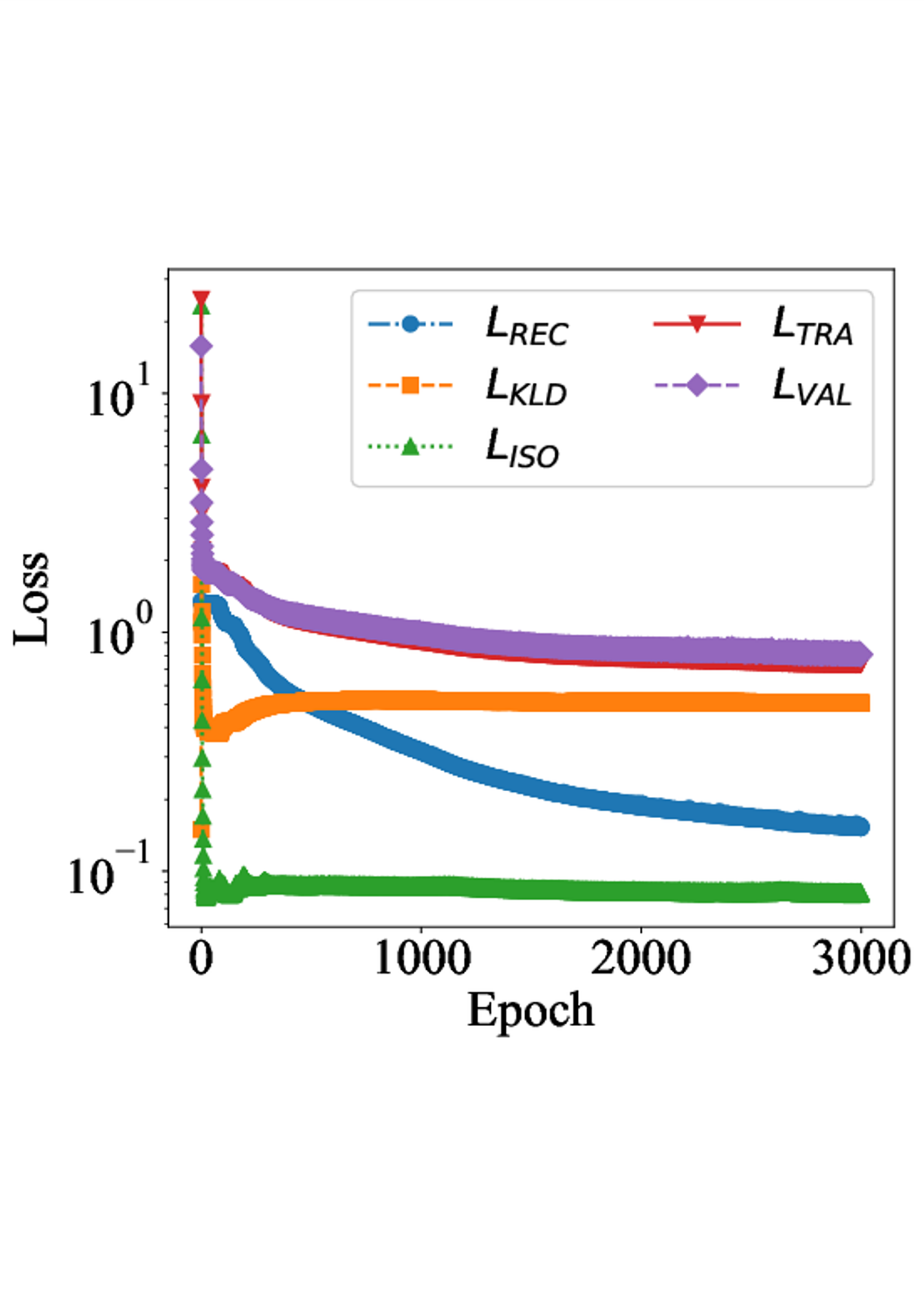}
        \caption{Loss}
        \label{fig:dnaloss}
    \end{subfigure}
    \hfill
    \begin{subfigure}{0.48\columnwidth}
        \includegraphics[width=\linewidth,trim=0 350 0 350,clip]{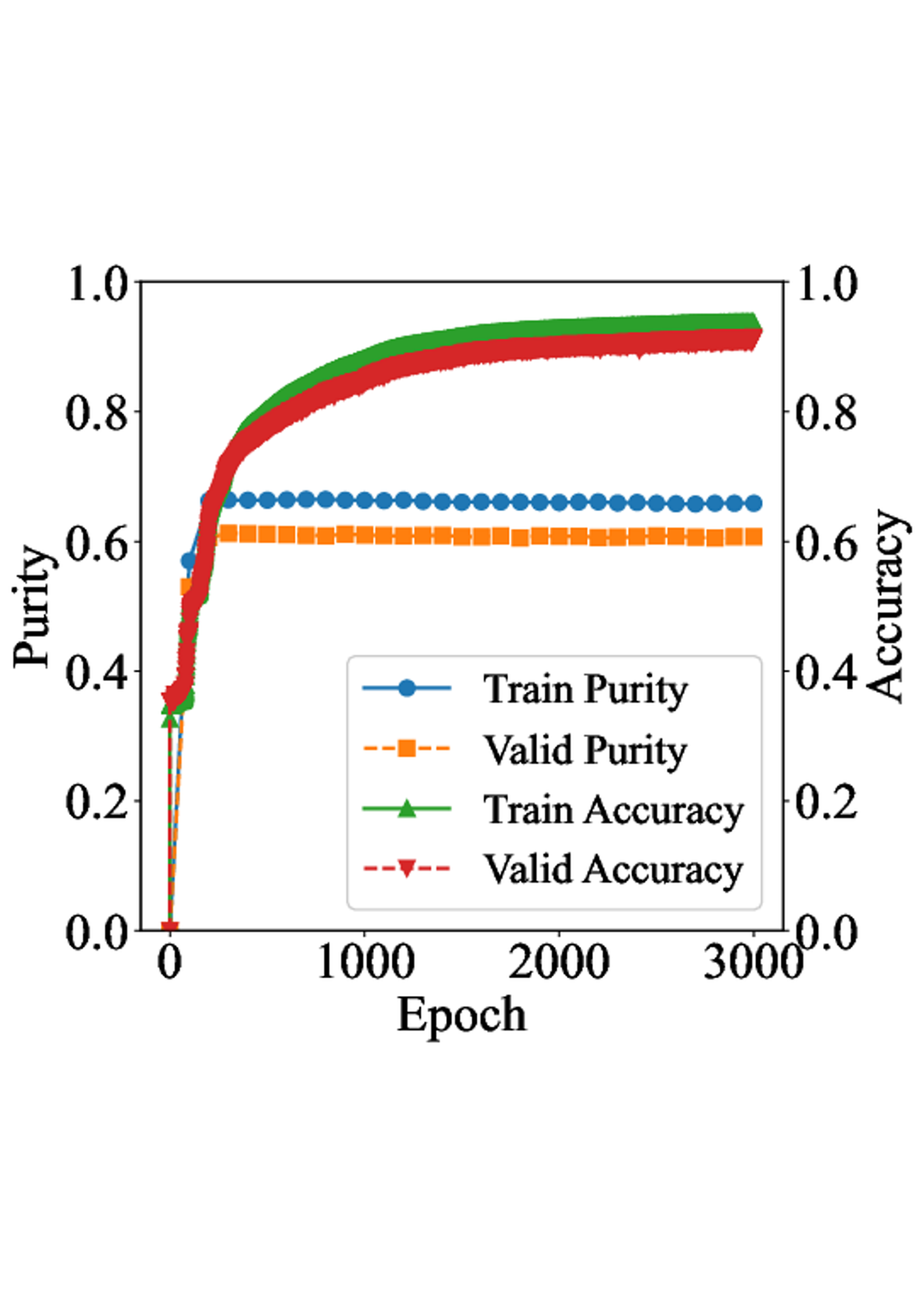}
        \caption{Accuracy and Purity}
        \label{fig:dnaaccuracy}
    \end{subfigure}

    \caption{Training performance of PrIVAE for Ag\textsubscript{N}-DNA design. (a) Loss components include reconstruction loss ($\mathcal{L}_{\mathrm{REC}}$), KL divergence ($\mathcal{L}_{\mathrm{KLD}}$), isometric regularization ($\mathcal{L}_{\mathrm{ISO}}$), total training loss ($\mathcal{L}_{\mathrm{TRA}}$), and validation loss ($\mathcal{L}_{\mathrm{VAL}}$). (b) Training and validation \(\text{Purity}_{15}\) (left axis) and accuracy (right axis).}
    \label{fig:DNA}\vsa
\end{figure}

\noindent{\bf Model performance.} The optimal PriVAE model for this data employs a bi-directional LSTM for sequence embedding; and graph smoothing and isometric regularization for organizing the latent space (alternative model results are reported in the ablation section and tuning hyperparameter ranges in the Appendix.) We report the training loss components as well as the overall training and validation loss as a function of training epochs in Fig.~\ref{fig:dnaloss}. The isometric loss quickly drops to a relatively low level, while the model takes several thousand epochs for the reconstruction loss to reduce to a comparable level. The corresponding purity and accuracy traces are presented in Fig.~\ref{fig:dnaaccuracy}. The reconstruction accuracy saturates at $94\%$ in training and $91\%$ in validation (effectively less than one of the $10$ DNA bases are incorrectly reconstructed on average). The \(\text{Purity}_{15}\) metrics saturate at $0.66$ for training and $0.61$ for validation, suggesting strong local alignment between spectral behavior and latent representations. 

\begin{figure}[t]
    \centering
    \includegraphics[width=0.6\columnwidth,trim={0 2.5cm 0 0.2cm},clip]{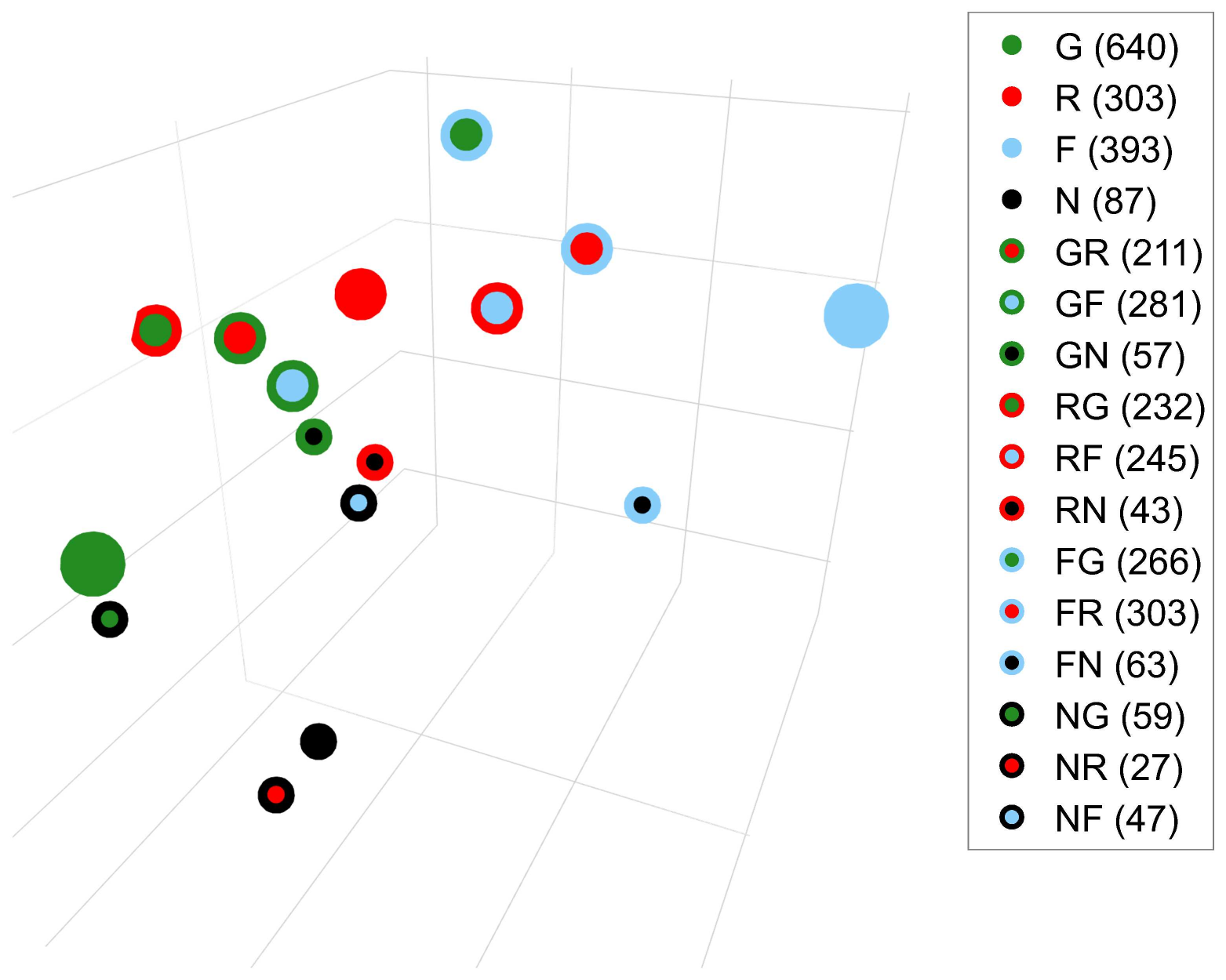}
    \caption{\footnotesize 3D PCA projection of the latent space for DNA-stabilized silver nanoclusters. Markers show the mean latent vector of a pseudo label, colored by strongest emission peak and marked with a secondary peak if present. Marker sizes reflect label frequencies and values in parentheses indicate training sample counts. 
    }\label{fig:DNA-latent}\vsa\vsa
\end{figure}

To further characterize the organization of the $22$-dimensional latent space $z$, we employ PCA (3D with $56\%$ retained variance) and visualize the means of groups of training instances with the same pseudo label in Fig.~\ref{fig:DNA-latent}. The outer color of markers indicates the primary peak color and the inner one (if present) designates the second brightest peak ($3$-rd and $4$-th brightest peaks are ignored for this visualization). 
The learned latent space organizes instances in meaningful subspaces according to the geometry of their properties $y$. Single-peak groups (G,R,F,N) occupy distinct regions, while dual-peaks instances lie between their respective counterparts. For example, the GF group is positioned between single-peak G and F and RN falls between R and N. This indicates that the model captures gradual transitions in emission properties. Notably, the G and N groups appear close in space, consistent with prior findings suggesting structural similarities between green- and NIR-emitting nanoclusters~\cite{sadeghi2024multi}.

\noindent{\bf De novo Ag\textsubscript{N}-DNA design.}  
Designing Near-Infrared (N) Ag\textsubscript{N}-DNAs is desired as NIR light penetrates biological tissues making the nanoclusters promising non-toxic molecular markers for deep tissue imaging~\cite{sadeghi2024multi}. At the same time designing new NIR sequences is challenging: only $7\%$ of our training sequences have a dominating NIR peak. Hence, we set out to design de novo NIR sequences by sampling in NIR-rich latent subspaces and decoding. To this end, we identify the $100$ training sequences of highest NIR purity (we re-define purity to consider only N peaks) to ensure that selected samples are surrounded by N-rich neighbors in latent space. We group them according to their labels (N, NG, NR, NF) and estimate means and covariance matrices for each group. We then sample $1000$ latent vectors from each of the regions and decode them to sequences. This targeted sampling strategy enables generation of candidate NIR-emitting sequences for downstream evaluation.

\begin{table}[tbp]
\footnotesize
\setlength{\tabcolsep}{2pt}
\begin{center}
\begin{tabular}{|c|c||c|c||c|c|} 
\hline
\footnotesize
  & {\bf Training} & \multicolumn{2}{c||}{\bf Sampling (1k per Grp)} & \multicolumn{2}{c|}{\bf Wet lab synth. (90 per Grp)} \\ \hline
\textbf{Grp} & {\bf \#Samp(\%)} & {\bf Resamp} & {\bf NIR Neigh}  & {\bf \% NIR (RC)} & {\bf \%  Dual (RC)}\\\hline
NG   & 59 (1.8)  & 7   &  23.5\% &  26 (14.2$\times$) & 10 (5.5$\times$)\\ \hline
NR   & 27 (0.8)  & 24  &  28.9\% & 13 (16.1$\times$) & 4.4 (5.4$\times$)\\\hline
NF   & 47 (1.4)  & 22  &  30.0\% &  19 (13.1$\times$) & 11 (7.7$\times$)\\\hline
N    & 87 (2.7)  & 21  &  23.7\% & 21 (7.9$\times$)  &  \\ \hline
\end{tabular}
\caption{\footnotesize Design of NIR DNA sequences with \ourmeth. We sample 1000 sequences from the high-purity regions for each label group with dominating (brightest) NIR peak (N,NG,NR,NF). \emph{Training:} Column \#Samp list the number of training samples in each group. \emph{Sampling:} Column Resamp lists the number of generated sequences identical to training samples (out of 1k). NIR Neigh is the average percentage of NIR-labeled nearest latent space neighbors of newly generated sequences (out of 15 NN). \emph{Wet lab synthesis:} Out of the 1000 candidates, we select the $90$ of highest NIR purity to synthesize and characterize spectrally. We quantify the success rate as the percentage of new sequences that have an NIR peak (\%NIR) and also that have dual peaks matching the target group (\%Dual). We also quantify the relative change (RC) of new sequences as the ratio of wet lab synthesis success rate (\%NIR or \%Dual)  and the corresponding abundance in training (Samp. \%).}
\label{tab:dnasampling}\vsa\vsa
\end{center}
\end{table}

Table~\ref{tab:dnasampling} lists statistics of sampling and experimental wet lab synthesis of candidates from the four sampling groups. It is important to note that the abundance of training instances (\#Samp) of the NIR groups is relatively low making this a challenging design problem. When sampling in latent space according to the group statistics, it is possible that upon decoding we obtain the same sequence as a training sample (Resamp column), though this happened at a low rate (less than 25 out of 1000 for all groups). Importantly, since we sample in high-purity regions, the neighborhoods of new samples are enriched in NIR peaks: between 23\% and 30\% of nearest neighbors of candidates have at least one NIR peak on average (column \emph{NIR Neigh}).

\noindent{\bf Wet lab synthesis.} We experimentally validate the top 90 predicted sequences (based on their NIR-Purity) for each of the four target groups. A sequence is an NIR success (column \% NIR) if it has peak in the near-infrared region, although it could also have peaks elsewhere. Dual-color success (column \% Dual) is the fraction of new sequences with emission peaks matching the target group (e.g., sampled from NG and producing a spectrum with both NIR and Green peaks). NIR success rates varied by group from 13\% for the NR group to as high as 26\% when sampling from the NG group.  
For dual-color emission, sampling in the NF group yielded the highest success rate, followed by the NG and NR groups. 

We also normalize the NIR (and Dual) success rates by training group frequency to quantify the level of enrichment of each of the groups (Relative change (RC) in the table). The increase in NIR emission was most pronounced when sampling from the NR group: 16.1 fold and lowest from the N group at 7.9 fold. In the dual-color emission, the best-performing group was NF at 7.7 fold, highlighting the model's capacity to restore its dual-emission property despite an imbalanced training distribution.

\subsection{Antimicrobial peptide design}
\noindent{\bf Data.} The GRAMPA dataset~\citep{witten2019deep} contains peptide sequences and their antimicrobial activity levels against multiple microorganisms. We select peptides of length between 7 and 20 amino acids representing approximately 57\% of all sequences and zero-pad sequences as needed to a fixed length of $20$. 
Each peptide is associated with a complex property $y$ quantifying its experimentally measured Minimum Inhibitory Concentration (MIC) values (low MIC stands for high antimicrobial activity) against three bacterial species: \textit{E. coli}, \textit{S. aureus}, and \textit{P. aeruginosa} (we select to work with these three as they are the most abundant in the dataset). We assign binary activity labels employing the threshold used for experimental validation in Szymczak et al.~\citep{szymczak2023discovering}. Namely, peptides with MIC $\leq$ 32\,\textmu g/mL (log MIC $\leq$ 1.51) are labeled as active (1), and those above the threshold as inactive (0) against a specific bacteria. We retain only peptides that have valid MIC measurements for at least two of the three bacterial species yielding a total of $2503$ peptides. Based on their activity profiles, peptides are categorized into 8 pseudo labels: N (inactive against all), E, S, P (active against one bacteria), ES, EP, SP (active against two), and ESP (active against all three). These pseudo labels are used solely for latent space visualization and purity analysis. We use the log-MIC profiles for model training and optimization.
\begin{figure}[t]
    \centering
    \begin{subfigure}[t]{0.48\columnwidth}
        \includegraphics[width=\linewidth,trim=0 350 0 350,clip]{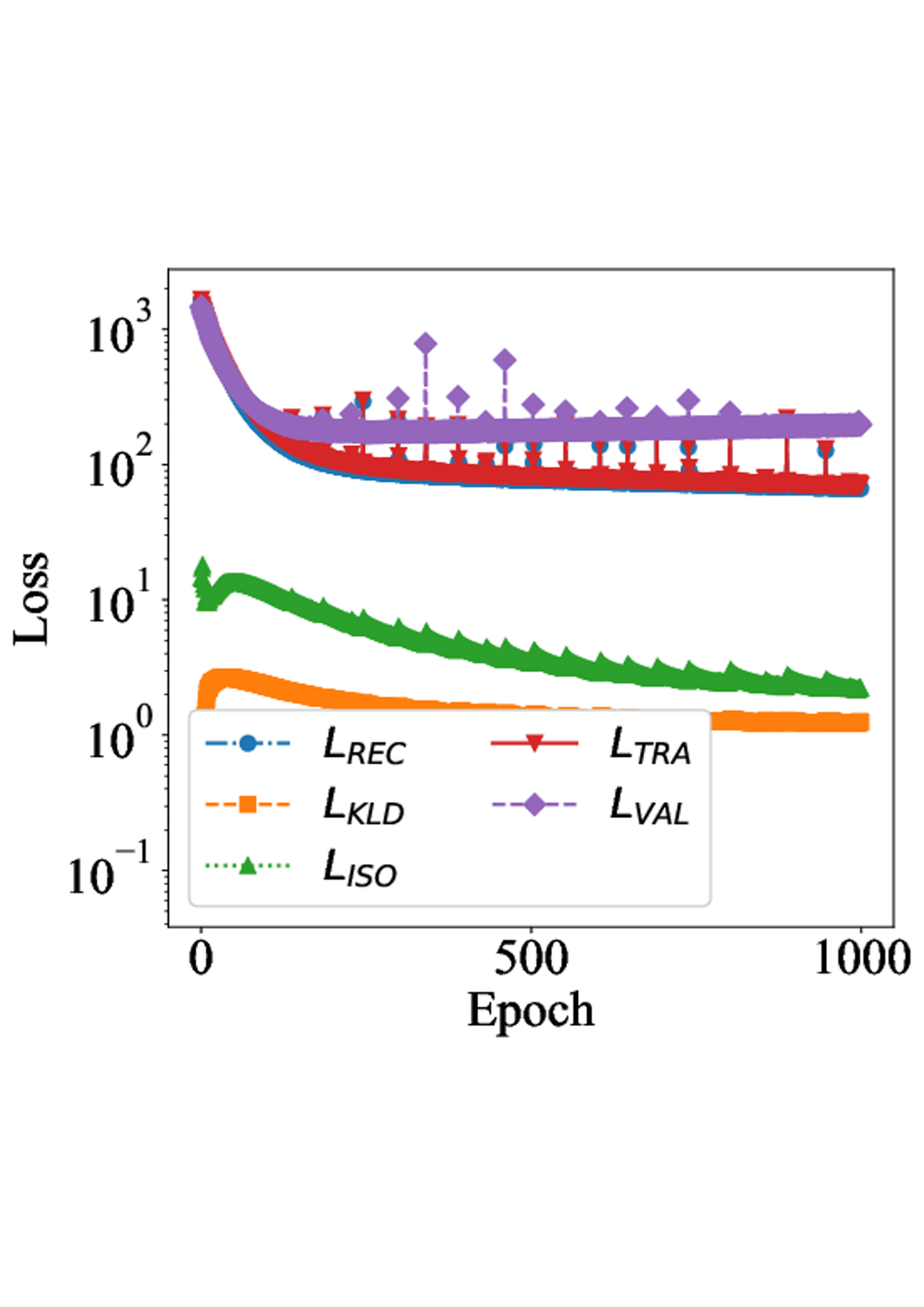}
        \caption{Loss}
        \label{fig:peploss}
    \end{subfigure}
    \hfill
    \begin{subfigure}[t]{0.48\columnwidth}
        \includegraphics[width=\linewidth,trim=0 350 0 350,clip]{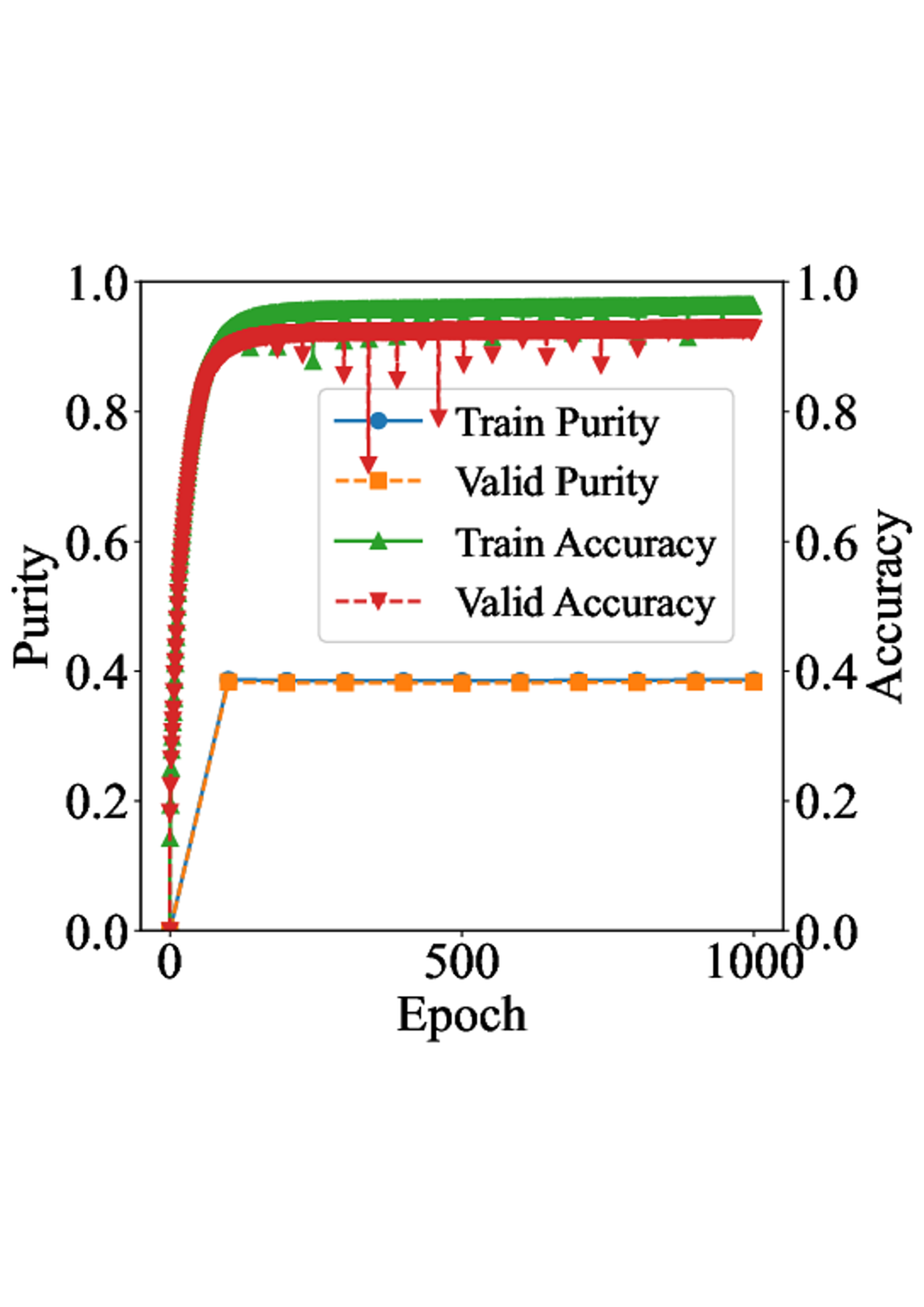}
        \caption{Accuracy and Purity}
        \label{fig:pepaccuracy}
    \end{subfigure}
    \caption{\footnotesize PrIVAE training on antimicrobial peptides. Training, validation and loss components (a) and accuracy and purity (b) as a function of training epochs. Note that the $L_{REC}$ curve is very close to that for $L_{TRA}$.}
    \label{fig:Pep}\vsa\vsa
\end{figure}

\begin{figure}[t]
    \centering
    \includegraphics[width=\linewidth,trim=0 600 0 600,clip]{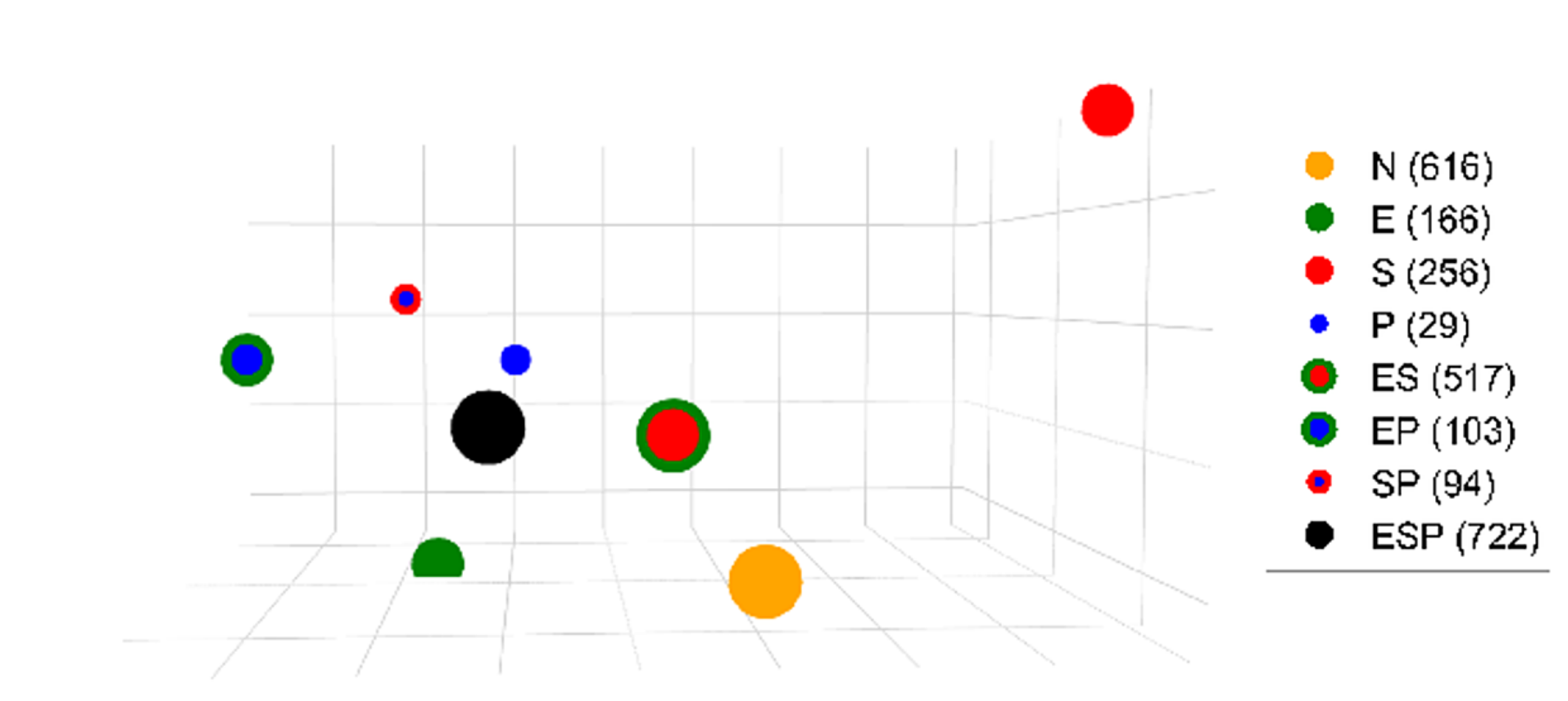}
    \caption{\footnotesize 3D PCA of the antimicrobial peptide latent space. Each point is the mean vector of a functional class, color-coded according to activity (e.g., ES is the group of peptides active against both E and S) and sized by pseudo label frequency. Sample counts are listed in parentheses in the legend.}
    \label{fig:Peplatent}
\end{figure}

\noindent{\bf Model performance.} We train \ourmeth for $1000$ epochs and exclude padding tokens from loss and accuracy calculations. Loss curves in Fig.~\ref{fig:peploss} converge across components, though the reconstruction loss dominates the overall training loss. The accuracy saturates at 96\% in training and 93\% in validation (Fig.~\ref{fig:pepaccuracy}), while the \(\text{Purity}_{15}\) metric reaches $0.39$ and $0.38$ for training and validation respectively.
We visualize the latent space similar to the DNA dataset in Fig.~\ref{fig:Peplatent}. Note, that in this case PCA retains only $40\%$ of the variance when reducing the $32$-dimensional latent space to 3D. While there is $60\%$ of variance that is not represented in this plot, trends in the the organization of the latent space are visible. Peptides active against multiple bacteria (e.g., EP, SP) appear between their corresponding single-activity groups, while the ESP (active against all) is positioned centrally among the active groups.

\begin{table}[tbp]
\scriptsize
\begin{center}
\renewcommand{\arraystretch}{1}
\setlength{\tabcolsep}{2.5pt}
\begin{tabular}{|c|c||c|c|c||c|c|c|}
\hline
\multirow{2}{*}{\textbf{Group}} & \multirow{2}{*}{\textbf{\#Samp}} 
& \multicolumn{3}{c||}{\textbf{PrIVAE}} 
& \multicolumn{3}{c|}{\textbf{Baseline VAE}} \\
\cline{3-8}
& & \textbf{F-Act.} & \textbf{P-Act.} & \textbf{Inact.} 
  & \textbf{F-Act.} & \textbf{P-Act.} & \textbf{Inact.} \\
\hline
E   & 166  & {\bf 87} &             & 13 & 67 &             & 33 \\ \hline
S   & 256  & {\bf 76} &             & 24 & 58 &             & 42 \\ \hline
P   & 29   & {\bf 78} &             & 22 & 50 &             & 50 \\\hline
ES  & 517  & 46 & 25 (20E,5S) & 29 & {\bf 54} & 24 (18E,6S) & 21 \\\hline
EP  & 103  & {\bf 60} & 28 (15E,13P)& 12 & 43 & 34 (34E,0P) & 23 \\\hline
SP  & 94   & {\bf 69} & 22 (6S,16P) & 9  & 19 & 62 (33S,29P)& 19 \\\hline
ESP & 722  & {\bf 44} & 48          & 8  & 6  & 86          & 8  \\
\hline
\end{tabular}
\caption{\footnotesize Oracle-based prediction results for generated peptides (100 sequences per group). \#Samp: number of training samples in each group. F-Act. (Fully Active): peptides predicted to be active against all intended strains; P-Act. (Partially Active): peptides predicted to be active against only one or a subset of the target strains; Inact.: predicted inactive against the group.}
\label{tab:oracle}\vsa\vsa
\end{center}
\end{table}

\noindent{\bf Peptide design.} Similar to our DNA design task, we sample near the ``purest'' latent training points from all active pseudo labels (i.e., excluding N). Since the model uses an LSTM decoder with fixed-length output, all generated peptides are of length $20$ amino acids. As with DNA, the generated $1000$ peptides per group were ranked based on their latent purity and the top $100$ candidates from each group were selected for downstream evaluation. To evaluate the antimicrobial activity of generated peptides, we employ the bacterial-specific activity prediction model based on AMP sequences from DBAASP~\cite{pirtskhalava2021dbaasp}. While not as reliable as direct wet lab experimentation, this model is a reasonable proxy oracle for success. It classifies peptides as active if their predicted MIC is $\leq 25\,\mu\mathrm{g}/\mathrm{ml}$. Since the model predicts single-bacteria activity at a time, designed broad-spectrum peptides (ESP, ES, EP and SP) were evaluated separately against each relevant microbe.

Results from the top $100$ highest-purity designed peptides per group  are presented in Tbl.~\ref{tab:oracle}. \emph{Fully Active (columns F-Act.)} are peptides active against all intended bacteria, while \emph{Partially Active} (columns P-Act.) lists the number of designed peptide active against a subset. Finally, the \emph{Inact.} column shows the number (out of 100) of designed peptides whiche were predicted inactive against any of the prescribed bacteria. Overall, PrIVAE outperformed the baseline VAE (no isometric regularization or graphs smoothing) in all groups except for ES. Even for groups with very limited training support like P and SP (both fewer than 100 training sequences) \ourmeth's top designed sequences are predicted to yield successful designs at very high rate: 78\% for P and 91\% for SP if counting partial success. Another important observation is the significantly higher predicted success of \ourmeth when designing for all 3 bacteria (ESP group): 44\% F-Act., compared to only 6\% with the basic VAE. The only group with higher success rate for the baseline was ES. Notably, this group is the second most frequent in training (517 sequences) which may have resulted in the baseline VAE over-representing its sequence motifs. In contrast, PrIVAE prioritizes property-geometry-aware representations giving it advantage with rarer properties which are arguably of highest interest for de novo design.

\subsection{Ablation analysis}

The central goal of our framework is to preserve the geometry of the property manifold in latent space and enable intuitive property-driven design by sampling. In this section we evaluate the contribution of individual components of \ourmeth (graph smoothing and isometric regularization) as well as the effect of architectural choices (LSTM vs attention-based sequence embedding).    
To assess the contribution of each geometry-preserving component, we compare four model variants: the baseline VAE (no GCN or isometric regularization), VAE+GCN which only incorporates graphs smoothing, VAE+Reg which incorporates isometric regularization no GCN, and \ourmeth combining both components. We also compare all these variants with attention-based sequence encoder and decoder instead of LSTM. We tune separately each model variant (hyperparameter ranges and tuning details are available in the Appendix) and report the best validation results (Pareto-optimal configurations considering both accuracy and purity). 

Table~\ref{tab:ablation_purity} summarizes the results for all model variants across both Ag\textsubscript{N}-DNA and peptide datasets, using LSTM and attention-based architectures. As expected, the baseline VAE achieves the highest reconstruction accuracy, since it is the least constrained and does not include any geometry-preserving mechanisms. We acknowledge that this near perfect accuracy, especially with short sequences, may suggest overfitting; however, it is expected, as the baseline model optimizes solely for reconstruction and can trivially memorize inputs. This comes at the cost of low purity in the latent space. The GCN component improves purity significantly, and the regularization-only model (VAE+Reg) helps with latent space organization to some extent. The combination of both graph smoothing and regularization in \ourmeth strikes the best balance between high purity and reconstruction accuracy in the 90\%-range. Although accuracy is modestly reduced compared to the unconstrained VAE, this reduction is desirable: the isometric regularizer encourages latent codes of sequences connected in the PNNG to remain close, preserving local geometry, while the GCN layers explicitly propagate and smooth hidden representations across PNNG neighbors through message passing. These constraints intentionally trade off perfect memorization for property-aware representations, and the resulting gain in purity demonstrates that this trade-off is necessary to prevent overfitting on short sequences and to enable generalizable, property-guided sequence design. While the performance of \ourmeth with LSTM and attention embeddings is close, the LSTM variant enabled higher validation purity in DNA design while maintaining high reconstruction accuracy ($>90\%$), motivating its use for de novo design and wet lab synthesis evaluation.
\begin{table}[tbp]
\footnotesize
\begin{center}
\setlength{\tabcolsep}{2.5pt}
\renewcommand{\arraystretch}{1}
\begin{tabular}{|c|c|c|c|c|c||c|c|c|c|}
\hline
& & \multicolumn{4}{|c||}{\textbf{LSTM}} & \multicolumn{4}{|c|}{\textbf{Attention}}  \\\hline
\multirow{2}{*}{\textbf{Data}} & \multirow{2}{*}{\textbf{Model}} & \multicolumn{2}{c|}{\textbf{Purity}} & \multicolumn{2}{c||}{\textbf{Accuracy}} & \multicolumn{2}{c|}{\textbf{Purity}} & \multicolumn{2}{c|}{\textbf{Accuracy}} \\
\cline{3-10}
& & \textbf{Val} & \textbf{Train} & \textbf{Val} & \textbf{Train} & \textbf{Val} & \textbf{Train} & \textbf{Val} & \textbf{Train} \\
\hline
\parbox[t]{2mm}{\multirow{4}{*}{\rotatebox[origin=c]{90}{Ag\textsubscript{N}-DNA}}}
& VAE & 0.40 & 0.39 & \textbf{0.99} & \textbf{1.00} & 0.28 & 0.27 & \textbf{1.00} & \textbf{1.00}\\
   & VAE+GCN & \underline{0.53} & \underline{0.63} & \underline{0.91} & \underline{0.94} & \underline{0.51} & \underline{0.60} & \underline{0.90} & \underline{0.93} \\
 & VAE+Reg & 0.41 & 0.40 & 0.86 & 0.88& 0.33 & 0.32 & 0.88 & 0.89  \\
 & \textbf{\ourmeth}& \textbf{0.61} & \textbf{0.66} & 0.90 & 0.93 & \textbf{0.54} & \textbf{0.68} & 0.76 & 0.80
 \\
\hline \hline
\parbox[t]{2mm}{\multirow{4}{*}{\rotatebox[origin=c]{90}{Peptides}}} & VAE & 0.25 & 0.26 & \textbf{0.93} & \textbf{1.00} & 0.26 & 0.26 & \textbf{0.97} & \textbf{1.00}\\
  & VAE+GCN & \underline{0.35} & \underline{0.36} & \textbf{0.93} & \underline{0.97} & \underline{0.34} & \underline{0.34} & 0.92 & \underline{0.96}\\
 & VAE+Reg & 0.29 & 0.29 & 0.92 & \textbf{1.00} & 0.27 & 0.27 & \underline{0.94} & \textbf{1.00}\\
 & \textbf{\ourmeth} & \textbf{0.38} & \textbf{0.39} & \textbf{0.93} & 0.96 & \textbf{0.38} & \textbf{0.39} & 0.92 & 0.95\\
\hline
\end{tabular}
\caption{\footnotesize Latent space purity and reconstruction accuracy for each model variant on Ag\textsubscript{N}-DNAs and peptide datasets. Models use either LSTM or multi-head attention for sequence processing, as indicated. Small standard deviations are omitted. Best results are \textbf{bold}-ed, while second-best \underline{underlined}.}
\label{tab:ablation_purity}\vsa\vsa\vsa
\end{center}
\end{table}

We evaluated the effect of replacing the Cauchy--Schwarz (CS) distance with alternative metrics. Many commonly used divergences, such as Kullback--Leibler, Bhattacharyya, Hellinger, or Wasserstein, do not admit closed-form solutions for Gaussian Mixture Models (GMMs) and therefore require approximation via sampling or numerical integration. Such approximations introduce stochasticity and additional scaling variability, which are undesirable in our setting where stability and reproducibility of property distances are important. In contrast, the CS divergence admits a closed-form expression for GMM-like structures, which motivated its use in the main experiments. To assess robustness, we additionally ablated CS using two simple closed-form metrics, $L_1$ (Manhattan) and $L_2$ (Euclidean) distances, computed directly on flattened GMM parameter vectors. Notably, the maximum observed CS distance in our dataset was $9.08$, whereas the maximum $L_1$ and $L_2$ distances reached $3{,}826{,}449.99$ and $3{,}156{,}211.46$, respectively, highlighting substantial scale disparities across distance functions. While the raw distance scale does not directly invalidate our method—since kernelization maps distances into $[0,1]$—extreme distance magnitudes cause kernel similarities to concentrate near zero. This effectively flattens the similarity distribution and weakens the discriminative structure of the property nearest-neighbor graph. Because the PNNG adjacency $W$ is used both in the isometric regularization term and in GNN-based smoothing, this loss of contrast in similarity values degrades neighborhood preservation in latent space. Consistent with this effect, replacing CS with $L_1$ or $L_2$ distances leads to a noticeable drop in latent purity across both RBF and Cauchy kernels, while classification accuracy remains largely unchanged (Table~\ref{tab:ablation_distance}).

A simple way to map distances $d$ to similarities $s$ is to normalize $d$ into $[0,1]$ using $d_{\text{norm}} = d / \max(d)$ and define $s = 1 - d_{\text{norm}}$. While intuitive, this linear mapping does not guarantee positive semi-definiteness (PSD) and may yield invalid kernel matrices. Instead, we consider kernels that are PSD under Euclidean distances, including the radial basis function (RBF) kernel,
$w_{ij} = \exp(-d(y_i,y_j)^2 / 2\sigma^2)$ with $\sigma=1$, and the Cauchy kernel,
$w_{ij} = 1 / (1 + d(y_i,y_j)^2)$. We also note that the Laplacian kernel,
$w_{ij} = \exp(-d(y_i,y_j)/\sigma)$, is closely related to the RBF kernel, differing primarily in its linear rather than quadratic decay with distance. Because both kernels induce similar locality behavior, we did not expect large qualitative differences and therefore focused our ablation on RBF and Cauchy kernels. The resulting purity and accuracy values under different distance–kernel combinations are reported in Table~\ref{tab:ablation_distance}, with all other hyperparameters held fixed.

\begin{table}[tbp]
\footnotesize
\begin{center}
\setlength{\tabcolsep}{3pt}
\renewcommand{\arraystretch}{1}
\begin{tabular}{|c|c||c|c||c|c|}
\hline
 &  & \multicolumn{4}{|c|}{\textbf{LSTM}} \\\hline
\multirow{2}{*}{\textbf{Data}} & \multirow{2}{*}{\textbf{PrIVAE Model}} & \multicolumn{2}{c||}{\textbf{Purity}} & \multicolumn{2}{c|}{\textbf{Accuracy}} \\
\cline{3-6}
 &  & \textbf{Val} & \textbf{Train} & \textbf{Val} & \textbf{Train} \\
\hline
\parbox[t]{2mm}{\multirow{6}{*}{\rotatebox[origin=c]{90}{Ag\textsubscript{N}-DNA}}}
& \textbf{CS \& RBF} & \textbf{0.61} & \textbf{0.66} & 0.90 & 0.93 \\
& L1 \& RBF       & 0.46 & 0.48 & \underline{0.91} & \underline{0.94} \\
& L2 \& RBF    &  0.45 & 0.46 & \textbf{0.92} & \underline{0.94} \\
& CS \& Cauchy & \underline{0.53} & \underline{0.63} & 0.88  & 0.93 \\
& L1 \& Cauchy       &  0.46 & 0.49 & \textbf{0.92} & \underline{0.94} \\
& L2 \& Cauchy    &  0.45 & 0.47 & \textbf{0.92} & \textbf{0.95} \\

\hline\hline
\parbox[t]{2mm}{\multirow{4}{*}{\rotatebox[origin=c]{90}{Peptides}}}
& L1 \& RBF  & \textbf{0.38} & \textbf{0.39} & \textbf{0.93} & \textbf{0.97} \\
& L2 \& RBF  &  \underline{0.37} & \textbf{0.39} &  \textbf{0.93} & \textbf{0.97}\\
& L1 \& Cauchy  & \underline{0.37} & \textbf{0.39} & \textbf{0.93} & \textbf{0.97} \\
& L2 \& Cauchy   & 0.36 & \textbf{0.39} & \underline{0.92} & \textbf{0.97} \\
\hline
\end{tabular}
\caption{\footnotesize
Ablation of distance functions and kernels used to construct the property nearest-neighbor graph (PNNG). Results are reported for different combinations of distance metrics (Cauchy--Schwarz, $L_1$, $L_2$) and similarity kernels (RBF, Cauchy) with all other model components and hyperparameters held fixed. While accuracy remains largely stable across choices, distance–kernel combinations that better preserve local property geometry (e.g., CS with RBF) yield higher latent purity, indicating improved neighborhood preservation in latent space. Values correspond to the last epoch. Best results are \textbf{bold}, second-best are \underline{underlined}.}

\label{tab:ablation_distance}
\end{center}
\end{table}

\section{Conclusion}
We presented \ourmeth, a generative framework for rational biological sequence design according to desired complex properties. Architecturally, \ourmeth is a geometry-preserving variational autoencoder that employs isometric regularization and property-nearest-neighbor message passing to align latent representations with functional property manifolds. It enabled controllable and interpretable sequence generation by targeted sampling from latent space regions enriched in the properties of interest. Applied to the design of DNA-stabilized silver nanoclusters and antimicrobial peptides, \ourmeth outperformed simpler baselines in reconstruction accuracy and latent space purity. We performed wet lab synthesis of predicted sequences for DNA nanoclusters and  discovered new (never synthesized before) nanostructures, resulting in up to a 16.1-fold enrichment of rare-property near-infrared emitters. Through ablation studies we confirmed that both geometric components are essential, highlighting PrIVAE’s principled and extensible value for property-guided biological sequence design across synthetic biological sequences, nanotechnology and beyond.

\input{arXiv_main.bbl}


\clearpage
\appendix
\section{Appendix}

\subsection{Source code, data and compute infrastructure}
All experiments were performed using PyTorch 2.0.0 and the Torch Geometric 2.5.3 libraries. The models were trained on a Dell compute server equipped with NVIDIA Tesla V100 GPUs (16GB memory per GPU). Training duration typically ranged from 2 to 5 hours depending on the dataset size, batch size, and model configuration. Code and data are available at \href{https://drive.google.com/drive/folders/1b4egKmzdscT5vnMuIk53E3rzYnFzET5E}{\url{https://drive.google.com/drive/folders/1b4egKmzdscT5vnMuIk53E3rzYnFzET5E}}.

\subsection{Distance measures for PNNG}
\label{append:dist}
To construct the PNNG, we need a distance measure that reflects functional similarity between high-dimensional property vectors. For the Ag\textsubscript{N}-DNA dataset, each DNA sequence is associated with an experimentally measured emission spectrum consisting of up to four peaks, each characterized by central wavelength $\lambda$, log-transformed intensity weight $v$, and peak width $\sigma$. While these spectra are not formal Gaussian Mixture Models (GMMs), their multi-peak structure is GMM-like for the purpose of defining pairwise distances. Following prior work~\cite{kampa2011closed}, we use the Cauchy-Schwarz (CS) divergence, which provides a closed-form and efficient means of comparing such structured profiles. The CS divergence between two distributions \( q(x) \) and \( p(x) \) is defined as:
\[
D_{CS}(q,p) = -\log\left(\frac{\int q(x)p(x)\,dx}{\sqrt{\int q(x)^2\,dx \int p(x)^2\,dx}}\right).
\]

Let
\[
q(x) = \sum_{m=1}^{M} \pi_m \mathcal{N}(x \mid \mu_m, \Lambda_m^{-1}), 
p(x) = \sum_{k=1}^{K} \tau_k \mathcal{N}(x \mid \nu_k, \Omega_k^{-1})
\]
where \(\pi_m, \mu_m, \Lambda_m\) and \(\tau_k, \nu_k, \Omega_k\) are the GMM weights, means, and precision matrices. Then, the closed-form expression for \(D_{CS}(q, p)\) is given by:

\begin{flalign*}
&D_{CS}(q,p) = -\log\left(\sum_{m=1}^{M}\sum_{k=1}^{K}\pi_{m}\tau_{k}z_{mk}\right) && \\
&+ \frac{1}{2}\log\left(\sum_{m=1}^{M}\frac{\pi_{m}^{2}|\Lambda_{m}|^{1/2}}{(2\pi)^{D/2}} + 2\sum_{m=1}^{M}\sum_{m'<m}\pi_{m}\pi_{m'}z_{mm'}\right) && \\
&+ \frac{1}{2}\log\left(\sum_{k=1}^{K}\frac{\tau_{k}^{2}|\Omega_{k}|^{1/2}}{(2\pi)^{D/2}} + 2\sum_{k=1}^{K}\sum_{k'<k}\tau_{k}\tau_{k'}z_{kk'}\right) &&
\end{flalign*}

where
\[
\begin{aligned}
z_{mk}  &= \mathcal{N}(\mu_m \mid \nu_k,\; \Lambda_m^{-1} + \Omega_k^{-1}), \\
z_{mm'} &= \mathcal{N}(\mu_m \mid \mu_{m'},\; \Lambda_m^{-1} + \Lambda_{m'}^{-1}), \\
z_{kk'} &= \mathcal{N}(\nu_k \mid \nu_{k'},\; \Omega_k^{-1} + \Omega_{k'}^{-1}).
\end{aligned}
\]

\balance

This expression enables exact and efficient computation of distances between spectra modeled in this parametric form, and critical for constructing the edge weights in PNNG. For full derivation, see Kampa et al.~\cite{kampa2011closed}.

For antimicrobial peptides, each peptide is associated with log-transformed MIC values against up to three bacterial strains. Since these are sparse, low-dimensional numeric vectors (as opposed to sets of peaks for the DNA data) we use the Manhattan (L1) distance to compare them. L1 distance is more robust to outliers than L2 (Euclidean) distance and provides a stable measure of dissimilarity when only a few values are present. 

It is important to note that the choice of distance function should be data-dependent, and different datasets may require different distance measures depending on the structure and characteristics of their associated properties.

\begin{table}[tbp]
\scriptsize
\begin{center}
\renewcommand{\arraystretch}{1}
\setlength{\tabcolsep}{3pt}
\begin{tabular}{|c|c||c|c||c|c|}
\hline
\multirow{2}{*}{\textbf{Parameter}} & \multirow{2}{*}{\textbf{Range}} 
& \multicolumn{2}{c||}{\textbf{Ag\textsubscript{N}-DNAs}} 
& \multicolumn{2}{c|}{\textbf{Peptides}} \\
\cline{3-6}
& & \textbf{LSTM} & \textbf{Attention} & \textbf{LSTM} & \textbf{Attention} \\
\hline
$\beta$ & [0.003, 0.01] & 0.007 & 0.007 & 0.007 & 0.007 \\
$\gamma$ & [0.2, 2.5] & 0.9132 & 0.836 & 0.589 & 1.02 \\
$h$  & [7, 128] & 11 & – & 16 & – \\
$k$  & [15, 30] & 17 & 17 & 25 & 20 \\
$|z|$  & [15, 512] & 22 & 20 & 32 & 32 \\
dropout & [0.0, 0.5] & 0 & 0 & 0 & 0 \\
head & [1, 6] & – & 2 & – & 2 \\
\hline
\end{tabular}
\caption{\footnotesize Search ranges and optimal hyperparameters selected by Optuna for LSTM- and attention-based models on Ag\textsubscript{N}-DNAs and Peptide datasets.}
\label{tab:hyperparam}
\end{center} \vsa\vsa
\end{table}
\subsection{Hyperparameter tuning and batching}
\label{append:param}
To identify well-performing model configurations that balance reconstruction accuracy, latent space purity, and isometric regularization, we employ \texttt{Optuna v4.0.0}, a modern hyperparameter optimization framework~\citep{akiba2019optuna}. Specifically, we use its \texttt{TPESampler} (Tree-structured Parzen Estimator), a Bayesian optimization algorithm that adaptively explores the search space. Unlike grid or random search, the \texttt{TPESampler} dynamically prioritizes promising regions based on previous trial outcomes, improving sample efficiency and convergence speed.

Table~\ref{tab:hyperparam} summarizes the hyperparameter search space and the final configurations selected by Optuna for each dataset and architecture. The KL divergence weight $\beta$ controls the strength of the variational regularization in the VAE loss. Higher values enforce stronger alignment with the prior but may reduce reconstruction accuracy, whereas lower values allow more flexibility at the cost of less-structured latents. The parameter $\gamma$ determines the weight of the isometric regularization term. Increasing $\gamma$ enforces geometric preservation more strictly. The PNNG neighbor count $k$ specifies how many property-nearest neighbors are used for graph construction. Smaller $k$ emphasizes local geometry while larger $k$ incorporates more global structure. The latent dimensionality $|z|$ defines the size of the learned latent space. For LSTM-based models, $h$ denotes the total output size of the bidirectional encoder, with each LSTM cell using a hidden size of $h/2$. For attention-based models, the number of attention heads is specified by the \texttt{head} parameter. The \texttt{dropout} value indicates the probability of zeroing activations during training for regularization. These parameters were tuned separately for each model variant to support a fair comparison under multi-objective optimization.

We formulate model selection as a multi-objective optimization problem: maximizing reconstruction accuracy and latent purity while minimizing isometric regularization. Optuna builds a Pareto front of optimal trade-offs, from which we select a configuration that balances all criteria and yields a functionally meaningful latent space.

Core minibatches for Ag\textsubscript{N}-DNAs were constructed by first grouping sequences according to their assigned spectral labels and sorting these labels in alphabetical order. Samples were then assembled into fixed-size batches of 32, prioritizing label homogeneity. If a label group did not contain enough sequences to complete a batch, the remaining slots were filled with unused sequences from other labels in the sorted list. This deterministic process was performed once prior to training and used consistently across all training epochs.

\end{document}

%% file: arXiv_main.bbl

%% file: arXiv_main.bbl
\begin{thebibliography}{10}
\providecommand{\url}[1]{#1}
\csname url@samestyle\endcsname
\providecommand{\newblock}{\relax}
\providecommand{\bibinfo}[2]{#2}
\providecommand{\BIBentrySTDinterwordspacing}{\spaceskip=0pt\relax}
\providecommand{\BIBentryALTinterwordstretchfactor}{4}
\providecommand{\BIBentryALTinterwordspacing}{\spaceskip=\fontdimen2\font plus
\BIBentryALTinterwordstretchfactor\fontdimen3\font minus \fontdimen4\font\relax}
\providecommand{\BIBforeignlanguage}[2]{{%
\expandafter\ifx\csname l@#1\endcsname\relax
\typeout{** WARNING: IEEEtran.bst: No hyphenation pattern has been}%
\typeout{** loaded for the language `#1'. Using the pattern for}%
\typeout{** the default language instead.}%
\else
\language=\csname l@#1\endcsname
\fi
#2}}
\providecommand{\BIBdecl}{\relax}
\BIBdecl

\bibitem{heinemann2006synthetic}
M.~Heinemann and S.~Panke, ``Synthetic biology—putting engineering into biology,'' \emph{Bioinformatics}, vol.~22, no.~22, pp. 2790--2799, 2006.

\bibitem{holmes2002novel}
T.~C. Holmes, ``Novel peptide-based biomaterial scaffolds for tissue engineering,'' \emph{Trends in biotechnology}, vol.~20, no.~1, pp. 16--21, 2002.

\bibitem{zhang2024artificial}
P.~Zhang, L.~Wei, J.~Li, and X.~Wang, ``Artificial intelligence-guided strategies for next-generation biological sequence design,'' \emph{Natl. Sci. Rev.}, vol.~11, no.~11, p. nwae343, 2024.

\bibitem{mardikoraem2023generative}
M.~Mardikoraem, Z.~Wang, N.~Pascual, and D.~Woldring, ``Generative models for protein sequence modeling: recent advances and future directions,'' \emph{Briefings in Bioinformatics}, vol.~24, no.~6, p. bbad358, 2023.

\bibitem{copp2019general}
S.~M. Copp, S.~M. Swasey, A.~Gorovits, P.~Bogdanov, and E.~G. Gwinn, ``General approach for machine learning-aided design of {DNA}-stabilized silver clusters,'' \emph{Chem. Mater.}, vol.~32, no.~1, pp. 430--437, 2019.

\bibitem{gonzalez2024atom}
A.~Gonz{\`a}lez-Rosell and S.~M. Copp, ``An atom-precise understanding of {DNA}-stabilized silver nanoclusters,'' \emph{Acc. Chem. Res.}, vol.~57, no.~15, pp. 2117--2129, 2024.

\bibitem{moomtaheen2022dna}
F.~Moomtaheen, M.~Killeen, J.~Oswald, A.~Gonz{\`a}lez-Rosell, P.~Mastracco, A.~Gorovits, S.~M. Copp, and P.~Bogdanov, ``{DNA}-stabilized silver nanocluster design via regularized variational autoencoders,'' in \emph{ACM SIGKDD conf. on Knowledge Discovery and Data Mining (KDD)}, 2022, pp. 3593--3602.

\bibitem{guha2023electron}
R.~Guha, A.~Gonz{\`a}lez-Rosell, M.~Rafik, N.~Arevalos, B.~B. Katz, and S.~M. Copp, ``Electron count and ligand composition influence the optical and chiroptical signatures of far-red and nir-emissive dna-stabilized silver nanoclusters,'' \emph{Chemical Science}, vol.~14, no.~41, pp. 11\,340--11\,350, 2023.

\bibitem{szymczak2023discovering}
P.~Szymczak, M.~Mo{\.z}ejko, T.~Grzegorzek, R.~Jurczak, M.~Bauer, D.~Neubauer, K.~Sikora, M.~Michalski, J.~Sroka, P.~Setny \emph{et~al.}, ``Discovering highly potent antimicrobial peptides with deep generative model {HydrAMP},'' \emph{Nat. Commun.}, vol.~14, no.~1, p. 1453, 2023.

\bibitem{greener2018design}
J.~G. Greener, L.~Moffat, and D.~T. Jones, ``Design of metalloproteins and novel protein folds using variational autoencoders,'' \emph{Scientific reports}, vol.~8, no.~1, p. 16189, 2018.

\bibitem{mastracco2022chemistry}
P.~Mastracco, A.~Gonz{\`a}lez-Rosell, J.~Evans, P.~Bogdanov, and S.~M. Copp, ``Chemistry-informed machine learning enables discovery of dna-stabilized silver nanoclusters with near-infrared fluorescence,'' \emph{ACS nano}, vol.~16, no.~10, pp. 16\,322--16\,331, 2022.

\bibitem{barazandeh2024utrgan}
S.~Barazandeh, ``{UTRGAN}: learning to generate 5' {UTR} sequences for optimized translation efficiency and gene expression,'' Master's thesis, Bilkent Universitesi (Turkey), 2024.

\bibitem{gupta2018feedback}
A.~Gupta and J.~Zou, ``Feedback gan (fbgan) for dna: a novel feedback-loop architecture for optimizing protein functions,'' \emph{arXiv preprint arXiv:1804.01694}, 2018.

\bibitem{yu2023multi}
H.~Yu, R.~Wang, J.~Qiao, and L.~Wei, ``Multi-cgan: deep generative model-based multiproperty antimicrobial peptide design,'' \emph{Journal of Chemical Information and Modeling}, vol.~64, no.~1, pp. 316--326, 2023.

\bibitem{soares2025targeted}
D.~Soares, L.~Hetzel, P.~Szymczak, F.~Theis, S.~G{\"u}nnemann, and E.~Szczurek, ``Targeted amp generation through controlled diffusion with efficient embeddings,'' \emph{arXiv preprint arXiv:2504.17247}, 2025.

\bibitem{torres2025generative}
M.~D. Torres, L.~T. Chen, F.~Wan, P.~Chatterjee, and C.~de~la Fuente-Nunez, ``Generative latent diffusion language modeling yields anti-infective synthetic peptides,'' \emph{Cell Biomater.}, 2025.

\bibitem{jin2025ampgen}
S.~Jin, Z.~Zeng, X.~Xiong, B.~Huang, L.~Tang, H.~Wang, X.~Ma, X.~Tang, G.~Shao, X.~Huang \emph{et~al.}, ``Ampgen: an evolutionary information-reserved and diffusion-driven generative model for de novo design of antimicrobial peptides,'' \emph{Communications Biology}, vol.~8, no.~1, p. 839, 2025.

\bibitem{jain2022biological}
M.~Jain, E.~Bengio, A.~Hernandez-Garcia, J.~Rector-Brooks, B.~F. Dossou, C.~A. Ekbote, J.~Fu, T.~Zhang, M.~Kilgour, D.~Zhang \emph{et~al.}, ``Biological sequence design with gflownets,'' in \emph{Intl. Conf. on Machine Learning (ICML)}.\hskip 1em plus 0.5em minus 0.4em\relax PMLR, 2022, pp. 9786--9801.

\bibitem{jain2023multi}
M.~Jain, S.~C. Raparthy, A.~Hern{\'a}ndez-Garc{\i}a, J.~Rector-Brooks, Y.~Bengio, S.~Miret, and E.~Bengio, ``Multi-objective gflownets,'' in \emph{International conference on machine learning}.\hskip 1em plus 0.5em minus 0.4em\relax PMLR, 2023, pp. 14\,631--14\,653.

\bibitem{zhu2024generative}
Y.~Zhu, Z.~Kong, J.~Wu, W.~Liu, Y.~Han, M.~Yin, H.~Xu, C.-Y. Hsieh, and T.~Hou, ``Generative ai for controllable protein sequence design: A survey,'' \emph{arXiv preprint arXiv:2402.10516}, 2024.

\bibitem{shcherbakova2024designing}
A.~Shcherbakova, D.~Buchan, and C.~P. Barnes, ``Designing minimal {E. coli} genomes using variational autoencoders,'' \emph{bioRxiv}, pp. 2024--10, 2024.

\bibitem{praljak2023protwave}
N.~Praljak, X.~Lian, R.~Ranganathan, and A.~L. Ferguson, ``Protwave-vae: Integrating autoregressive sampling with latent-based inference for data-driven protein design,'' \emph{ACS synthetic biology}, vol.~12, no.~12, pp. 3544--3561, 2023.

\bibitem{ozden2023rnagen}
F.~Ozden, S.~Barazandeh, D.~Akboga, S.~S. Tabrizi, U.~O.~S. Seker, and A.~E. Cicek, ``Rnagen: A generative adversarial network-based model to generate synthetic rna sequences to target proteins,'' \emph{bioRxiv}, pp. 2023--07, 2023.

\bibitem{chiquitto2024generative}
A.~G. Chiquitto, L.~S. Oliveira, P.~H. Bugatti, P.~T.~M. Saito, M.~Basham, R.~T. Raittz, and A.~R. Paschoal, ``Generative approaches for nucleotide sequences to enhance non-coding rna classification,'' \emph{bioRxiv}, pp. 2024--11, 2024.

\bibitem{andress2023daptev}
C.~Andress, K.~Kappel, M.~E. Villena, M.~Cuperlovic-Culf, H.~Yan, and Y.~Li, ``Daptev: Deep aptamer evolutionary modelling for covid-19 drug design,'' \emph{PLOS Computational Biology}, vol.~19, no.~7, p. e1010774, 2023.

\bibitem{zhang2024rnagenesis}
Z.~Zhang, L.~Chao, R.~Jin, Y.~Zhang, G.~Zhou, Y.~Yang, Y.~Yang, K.~Huang, Q.~Yang, Z.~Xu \emph{et~al.}, ``Rnagenesis: Foundation model for enhanced rna sequence generation and structural insights,'' \emph{bioRxiv}, pp. 2024--12, 2024.

\bibitem{madani2023large}
A.~Madani, B.~Krause, E.~R. Greene, S.~Subramanian, B.~P. Mohr, J.~M. Holton, J.~L. Olmos, C.~Xiong, Z.~Z. Sun, R.~Socher \emph{et~al.}, ``Large language models generate functional protein sequences across diverse families,'' \emph{Nat Biotechnol}, vol.~41, no.~8, pp. 1099--1106, 2023.

\bibitem{li2024discdiff}
Z.~Li, Y.~Ni, W.~A. Beardall, G.~Xia, A.~Das, G.-B. Stan, and Y.~Zhao, ``Discdiff: Latent diffusion model for dna sequence generation,'' \emph{arXiv preprint arXiv:2402.06079}, 2024.

\bibitem{huang2024latent}
K.~Huang, Y.~Yang, K.~Fu, Y.~Chu, L.~Cong, and M.~Wang, ``Latent diffusion models for controllable rna sequence generation,'' \emph{arXiv preprint arXiv:2409.09828}, 2024.

\bibitem{dasilva2024dna}
L.~F. DaSilva, S.~Senan, Z.~M. Patel, A.~J. Reddy, S.~Gabbita, Z.~Nussbaum, C.~M.~V. C{\'o}rdova, A.~Wenteler, N.~Weber, T.~M. Tunjic \emph{et~al.}, ``Dna-diffusion: leveraging generative models for controlling chromatin accessibility and gene expression via synthetic regulatory elements,'' \emph{bioRxiv}, 2024.

\bibitem{lu2024cell}
S.~Z. Lu, Z.~Lu, E.~Hajiramezanali, T.~Biancalani, Y.~Bengio, G.~Scalia, and M.~Koziarski, ``Cell morphology-guided small molecule generation with gflownets,'' in \emph{ICML 2024 Workshop on Structured Probabilistic Inference $\{$$\backslash$\&$\}$ Generative Modeling}, 2024.

\bibitem{dean2020variational}
S.~N. Dean and S.~A. Walper, ``Variational autoencoder for generation of antimicrobial peptides,'' \emph{ACS omega}, vol.~5, no.~33, pp. 20\,746--20\,754, 2020.

\bibitem{dean2021pepvae}
S.~N. Dean, J.~A.~E. Alvarez, D.~Zabetakis, S.~A. Walper, and A.~P. Malanoski, ``Pepvae: variational autoencoder framework for antimicrobial peptide generation and activity prediction,'' \emph{Frontiers in microbiology}, vol.~12, p. 725727, 2021.

\bibitem{das2018pepcvae}
P.~Das, K.~Wadhawan, O.~Chang, T.~Sercu, C.~D. Santos, M.~Riemer, V.~Chenthamarakshan, I.~Padhi, and A.~Mojsilovic, ``Pepcvae: Semi-supervised targeted design of antimicrobial peptide sequences,'' \emph{arXiv preprint arXiv:1810.07743}, 2018.

\bibitem{surana2023pandoragan}
S.~Surana, P.~Arora, D.~Singh, D.~Sahasrabuddhe, and J.~Valadi, ``Pandoragan: generating antiviral peptides using generative adversarial network,'' \emph{SN Computer Science}, vol.~4, no.~5, p. 607, 2023.

\bibitem{sadeghi2024multi}
E.~Sadeghi, P.~Mastracco, A.~Gonz{\`a}lez-Rosell, S.~M. Copp, and P.~Bogdanov, ``Multi-objective design of dna-stabilized nanoclusters using variational autoencoders with automatic feature extraction,'' \emph{ACS nano}, vol.~18, no.~39, pp. 26\,997--27\,008, 2024.

\bibitem{schmidinger2024bio}
N.~Schmidinger, L.~Schneckenreiter, P.~Seidl, J.~Schimunek, P.-J. Hoedt, J.~Brandstetter, A.~Mayr, S.~Luukkonen, S.~Hochreiter, and G.~Klambauer, ``Bio-xlstm: Generative modeling, representation and in-context learning of biological and chemical sequences,'' \emph{arXiv preprint arXiv:2411.04165}, 2024.

\bibitem{duque2022geometry}
A.~F. Duque, S.~Morin, G.~Wolf, and K.~R. Moon, ``Geometry regularized autoencoders,'' \emph{IEEE Trans. Pattern Anal. Mach. Intell.}, vol.~45, no.~6, pp. 7381--7394, 2022.

\bibitem{lee2022regularized}
Y.~Lee, S.~Yoon, M.~Son, and F.~C. Park, ``Regularized autoencoders for isometric representation learning,'' in \emph{Intl. Conf. on Learning Representations (ICLR)}, 2022.

\bibitem{palmaenforcing}
A.~Palma, S.~Rybakov, L.~Hetzel, S.~G{\"u}nnemann, and F.~J. Theis, ``Enforcing latent euclidean geometry in vaes for statistical manifold interpolation.''

\bibitem{xu2024beyond}
J.~Xu, A.~Moskalev, T.~Mansi, M.~Prakash, and R.~Liao, ``Beyond sequence: Impact of geometric context for rna property prediction,'' \emph{arXiv preprint arXiv:2410.11933}, 2024.

\bibitem{krapp2024context}
L.~F. Krapp, F.~A. Meireles, L.~A. Abriata, J.~Devillard, S.~Vacle, M.~J. Marcaida, and M.~Dal~Peraro, ``Context-aware geometric deep learning for protein sequence design,'' \emph{Nature Communications}, vol.~15, no.~1, p. 6273, 2024.

\bibitem{zhu2023geometric}
J.~Zhu, J.~H. Oh, A.~K. Simhal, R.~Elkin, L.~Norton, J.~O. Deasy, and A.~Tannenbaum, ``Geometric graph neural networks on multi-omics data to predict cancer survival outcomes,'' \emph{Computers in biology and medicine}, vol. 163, p. 107117, 2023.

\bibitem{higgins2017beta}
I.~Higgins, L.~Matthey, A.~Pal, C.~Burgess, X.~Glorot, M.~Botvinick, S.~Mohamed, and A.~Lerchner, ``beta-vae: Learning basic visual concepts with a constrained variational framework,'' in \emph{Intl. Conf. on Learning Representations (ICLR)}, 2017.

\bibitem{lin2023evolutionary}
Z.~Lin, H.~Akin, R.~Rao, B.~Hie, Z.~Zhu, W.~Lu, N.~Smetanin, R.~Verkuil, O.~Kabeli, Y.~Shmueli \emph{et~al.}, ``Evolutionary-scale prediction of atomic-level protein structure with a language model,'' \emph{Science}, vol. 379, no. 6637, pp. 1123--1130, 2023.

\bibitem{tenenbaum2000global}
J.~B. Tenenbaum, V.~d. Silva, and J.~C. Langford, ``A global geometric framework for nonlinear dimensionality reduction,'' \emph{Science}, vol. 290, no. 5500, pp. 2319--2323, 2000.

\bibitem{roweis2000nonlinear}
S.~T. Roweis and L.~K. Saul, ``Nonlinear dimensionality reduction by locally linear embedding,'' \emph{Science}, vol. 290, no. 5500, pp. 2323--2326, 2000.

\bibitem{belkin2003laplacian}
M.~Belkin and P.~Niyogi, ``Laplacian eigenmaps for dimensionality reduction and data representation,'' \emph{Neural Comput.}, vol.~15, no.~6, pp. 1373--1396, 2003.

\bibitem{kampa2011closed}
K.~Kampa, E.~Hasanbelliu, and J.~C. Principe, ``Closed-form cauchy-schwarz pdf divergence for mixture of gaussians,'' in \emph{Intl. Joint Conf. on Neural Networks (IJCNN)}.\hskip 1em plus 0.5em minus 0.4em\relax IEEE, 2011, pp. 2578--2585.

\bibitem{meilua2024manifold}
M.~Meil{\u{a}} and H.~Zhang, ``Manifold learning: What, how, and why,'' \emph{Annual Review of Statistics and Its Application}, vol.~11, no.~1, pp. 393--417, 2024.

\bibitem{cayton2008algorithms}
L.~Cayton \emph{et~al.}, \emph{Algorithms for manifold learning}.\hskip 1em plus 0.5em minus 0.4em\relax eScholarship, University of California, 2008.

\bibitem{lim2024graph}
J.~Lim, J.~Kim, Y.~Lee, C.~Jang, and F.~C. Park, ``Graph geometry-preserving autoencoders,'' in \emph{Forty-first International Conference on Machine Learning}, 2024.

\bibitem{gilmer2017neural}
J.~Gilmer, S.~S. Schoenholz, P.~F. Riley, O.~Vinyals, and G.~E. Dahl, ``Neural message passing for quantum chemistry,'' in \emph{International conference on machine learning}.\hskip 1em plus 0.5em minus 0.4em\relax PMLR, 2017, pp. 1263--1272.

\bibitem{kipf2016semi}
T.~N. Kipf and M.~Welling, ``Semi-supervised classification with graph convolutional networks,'' \emph{arXiv preprint arXiv:1609.02907}, 2016.

\bibitem{velivckovic2017graph}
P.~Veli{\v{c}}kovi{\'c}, G.~Cucurull, A.~Casanova, A.~Romero, P.~Lio, and Y.~Bengio, ``Graph attention networks,'' \emph{arXiv preprint arXiv:1710.10903}, 2017.

\bibitem{witten2019deep}
J.~Witten and Z.~Witten, ``Deep learning regression model for antimicrobial peptide design,'' \emph{BioRxiv}, p. 692681, 2019.

\bibitem{copp2021large}
S.~M. Copp and A.~Gonz{\`a}lez-Rosell, ``Large-scale investigation of the effects of nucleobase sequence on fluorescence excitation and stokes shifts of dna-stabilized silver clusters,'' \emph{Nanoscale}, vol.~13, no.~8, pp. 4602--4613, 2021.

\bibitem{pirtskhalava2021dbaasp}
M.~Pirtskhalava, A.~A. Amstrong, M.~Grigolava, M.~Chubinidze, E.~Alimbarashvili, B.~Vishnepolsky, A.~Gabrielian, A.~Rosenthal, D.~E. Hurt, and M.~Tartakovsky, ``Dbaasp v3: database of antimicrobial/cytotoxic activity and structure of peptides as a resource for development of new therapeutics,'' \emph{Nucleic Acids Res.}, vol.~49, no.~D1, pp. D288--D297, 2021.

\bibitem{akiba2019optuna}
T.~Akiba, S.~Sano, T.~Yanase, T.~Ohta, and M.~Koyama, ``Optuna: A next-generation hyperparameter optimization framework,'' in \emph{Proceedings of the 25th ACM SIGKDD international conference on knowledge discovery \& data mining}, 2019, pp. 2623--2631.

\end{thebibliography}
